\documentclass[12pt]{article}
\pdfoutput=1 % if your are submitting a pdflatex (i.e. if you have
\usepackage[utf8]{inputenc}
\usepackage{indentfirst}
\usepackage{amsmath}
\usepackage{graphicx}
\usepackage{caption}
\usepackage{subcaption}
\usepackage{times}
\usepackage{authblk}

\usepackage{amssymb}
\usepackage{amsbsy}
\usepackage{mathtools}
\usepackage[left, running, mathlines]{lineno}
 \usepackage{mathptmx}           % this gives slanted Times symbols for math
\usepackage{bm}   % for bold math fonts, especially greek
\usepackage{framed}  % for frame boxes
\numberwithin{equation}{section}
\newcommand{\bit}{\begin{itemize}}
\newcommand{\eit}{\end{itemize}}

\def\mtx#1{\mathsf{#1}}

\def\benu{\begin{enumerate}}
\def\eenu{\end{enumerate}}
\def\noi{\noindent}
\def\btab{\begin{tabbing}}
\def\etab{\end{tabbing}}

\def\bit{\begin{itemize}}
\def\eit{\end{itemize}}
\def\beq{\begin{equation}}
\def\eeq{\end{equation}}
\def\bec{\begin{center}}
\def\eec{\end{center}}
\def\btable{\begin{tabular}}
\def\etable{\end{tabular}}
\def\beqr{\begin{eqnarray}}
\def\eeqr{\end{eqnarray}}

\def\Rarw{\Rightarrow}

\def\gm{\gamma}

\def\lm{\lambda}
\def\eps{\epsilon}

\def\al{\alpha}
\def\bt{\beta}

\def\dl{\delta}
\def\Dl{\Delta}
\def\sg{\sigma}

\def\del{\partial}

\def\half{\frac{1}{2}}

\def\btab{\begin{tabbing}}
\def\etab{\end{tabbing}}
\def\beqrs{\begin{eqnarray*}}
\def\eeqrs{\end{eqnarray*}}
\def\noi{\noindent}
\def\lan{\langle}
\def\ran{\rangle}
\def\bibi{\bibitem}
\def\bfig{\begin{figure}}
\def\efig{\end{figure}}
\def\fr{\frac}

\def\barr{\begin{array}}
\def\earr{\end{array}}

 \fontfamily{ptm}\selectfont
\title{\mbox{}  \hfill  {\normalsize FERMILAB-TM-2753-AD } \\   \mbox{} \\
  Modeling Transverse Space Charge effects in IOTA with pyORBIT}

\author[2]{Runze Li}
\author[1]{Tanaji Sen  \footnote{tsen@fnal.gov}}
\author[1]{Jean-Francois Ostiguy}

\affil[1]{Fermi National Accelerator Laboratory, Batavia, IL 60510 }
\affil[2]{ University of Wisconsin, Madison, WI 53706 }

 \date{}

\begin{document}
\maketitle
\begin{abstract}
  The role and mitigation of space charge effects are important aspects of the beam physics research to be
  performed at the Integrable Optics Test
  Accelerator (IOTA) at Fermilab. The impact of nonlinear integrability (partial and complete) on  space charge
  driven incoherent and coherent resonances will be a new feature of this accelerator and has the potential to
  influence the design of future low energy proton synchrotrons. In this report we model transverse space charge
  effects using the PIC code
  pyORBIT. First we benchmark the single particle tracking results against MADX with checks on symplecticity, tune
  footprints,  and dynamic aperture in a partially integrable lattice realized with a special octupole string insert.
  Our space charge calculations begin with an examination of the  4D symplecticity. Short term tracking is done first with
  the initial bare lattice and then with a  matching of the  rms values with space charge. 
Next, we  explore slow initialization of charge as a
  technique to establish steady state and reduce emittance growth and beam loss following injection into a
  mismatched lattice. We establish values of space charge simulation parameters so as to ensure numerical
  convergence. Finally, we compare the simulated space charge induced tune shifts and footprints against theory. 
\end{abstract}
\clearpage
\tableofcontents
\clearpage
\section{Introduction}
 The Fermilab Integrable Optics Test Accelerator (IOTA) is a storage ring for beam physics research.
 An important aspect of the research program is to explore the potential of integrable optics to mitigate deleterious
 effects   of space charge in high intensity proton synchrotrons. In theory, integrable single particle dynamics eliminate resonances,
  providing stable motion over a wide tune range. In a beam, large betatron tune spread is known to be effective at suppressing instabilities. To the extent that desirable properties of integrable single particle motion persist in the presence of strong space charge, integrability may help to preserve stability in intense beams, but that remains to be verified.

The IOTA ring is currently operating with electrons in the energy range 100 - 150 MeV.   At a later stage, experiments
will be performed with protons at a kinetic energy of 2.5 MeV. 
 By turning on/off special nonlinear inserts, the ring can be operated with either conventional 
  or integrable optics.  Space charge in the proton beam will be  a determinant factor for beam stability. As in any other low energy
  proton synchrotron, both incoherent and coherent space charge effects will play a role. When the ring is configured with
  integrable  optics,  the impact of the perturbation due to space charge on properties associated with integrability needs to be understood. 
 In this context, it is important to  assess the capability and suitability of existing simulation codes.
 We report here on a variety of relevant validation tests that were
 performed using pyORBIT \cite{pyORBIT}, a PIC code developed and maintained at ORNL and motivated by the need to
 simulate  certain aspects of SNS.       
\begin{table}
  \bec
  \btable{|c|c|} \hline
  \multicolumn{2}{|c||}{IOTA proton parameters} \\ \hline
  Circumference & 39.97 [m] \\
  Kinetic Energy & 2.5 [MeV] \\
Maximum bunch intensity /current & 9$\times 10^{10}$ / 8 [mA] \\
Transverse normalized rms emittance & (0.3, 0.3) [mm-mrad] \\
Betatron tunes & (5.3, 5.3) \\
Natural chromaticities &  (-8.2, -8.1)     \\ 
Average transverse beam sizes (rms) &  (2.22, 2.22) [mm] \\ 
Kinematic $\gm$ / Transition $\gm_T$ &  1.003 / 3.75 \\
Rf voltage & 400 [V]  \\
Rf frequency  /  harmonic number & 2.2 [MHz]   / 4   \\
Bucket  wavelength &  $\sim 10$ [m] \\
Bucket half height in $\Dl p/p$ &   3.72 $\times 10^{-3}$ \\
rms bunch length &  1.7 [m] \\
rms energy /momentum spread &  1.05$\times 10^{-5}$ / 1.99 $\times 10^{-3}$   \\
Beam pipe radius & 25 [mm] \\
Bunch density &    6.9 $\times 10^{14}$ [m$^{-3}$] \\
Plasma period $\tau_p$ &  0.18 [$\mu$-sec] /  0.1 [turns] \\
Average Debye length $\lm_D$  &  559 [$\mu$m] \\
\hline
\etable
\caption{Machine and beam parameters of the IOTA proton ring}
\label{table:parameters}
\eec
\end{table}

 Parameters for the IOTA proton ring are presented in Table \ref{table:parameters}. The plasma period is included because it is a relevant time scale for initial emittance growth \cite{Wangler_91} mechanisms. For example, charge redistribution has the shortest time scale
$\tau_p/4$ while rms mismatch or energy transfer between planes have a time scale of $\sim 10 \tau_p$.
Taking into account relativistic corrections, the plasma period is \cite{Reiser}:
\beq
\tau_p = 2 \pi \sqrt{\fr{\eps_0 m_0 \gm^3 }{ n  e^2 }}
\eeq
Here $\eps_0$ is the vacuum permittivity, $m_0$ is the proton mass and $n$ is the number density of the
proton bunch. The volume of a Gaussian bunch with extent $\pm (3\sg_x, 3 \sg_y, 3\sg_z)$ along the three axes
is
\beq
V = \sg_x \sg_y \sg_z \left( \int_{-3}^{3} \exp[- \fr{u^2}{2}] du \right)^3
= (2 \pi )^{3/2}[\mathrm{erf}(3/\sqrt{2})]^3 \sg_x \sg_y \sg_z
\eeq
Substituting $\sg_x = \sg_y = 2.2 \;\text{mm}$ and $\sg_z = 1.7$ m we obtain $\tau_p= 0.1$ turn, which  suggests
that initial emittance growth will occur on the scale of a single turn. 

The Debye length $\lm_D$ is \cite{Reiser}
\beq
\lm_D = \fr{\sqrt{\lan v_{\perp}^2 \ran} \tau_p }{2\pi} = \sqrt{\fr{ \eps_0 \gm^2 k_B T_{\perp}}{e^2 n} }
\eeq
The  temperature in the horizontal plane is 
\beq
k_B T_{x} = m_0 \gm  \lan v_{x}^2 \ran = m_0 \gm c^2 (\gm^2 - 1)   \lan (x')^2 \ran = m_0 \gm c^2 (\gm^2 - 1)
\gm_x \eps_x
\eeq
where $\gm_x$ is the lattice Twiss function and $\eps_x$ is the un-normalized emittance. 
Averaging this over the whole ring, we find the average temperature over the ring (see Appendix A)
\beq
\lan k_B T_x \ran = m_0 c^2 (\gm^2 - 1) \eps_{x, N} [\fr{-2 Q_x'}{R}]
  \eeq
  where $Q_x'$ is the natural horizontal chromaticity, $R$ is the machine radius and $\eps_{x, N}$ is the
  normalized emittance. This relation shows for example that the 
average transverse temperature in a circular machine depends on  the natural chromaticity and ring radius.
Substituting, we find that the average Debye length over the ring is
  \beq
  \lan \lm_D \ran = \sqrt{\left[ \fr{ \eps_0  m_0 c^2 \gm^2 (\gm^2 - 1)\eps_{\perp, N}}{e^2 n} \right]
 \left[  \fr{(-2 Q_{\perp}')}{ R} \right] }
  \eeq
  The first square bracket in this equation contains beam dependent variables while the second bracket
  isolates the machine dependent quantities. 
  The Debye length sets the scale for the importance of space charge effects;  the regime $\lan \lm_D \ran \ll$ beam radius is  space charge dominated.  Table \ref{table:parameters} shows that with an average beam size of 2.2 mm and $\lm_D = 0.56$ mm, IOTA will be in the space charge dominated regime at full intensity.
\section{Single-Particle Tracking}
In this section, we aim to validate single-particle tracking in pyORBIT against MADX, a well-established and extensively tested code.
To minimize discrepancies possibly introduced by subtle differences in the way distributed nonlinearities are modeled by the two codes, all sources of nonlinearity other than those confined to a special insert region are turned off.  The insert region is initially populated with octupole magnets whose strengths vary in inverse proportion to $\beta^3(s)$; theoretically, the dynamics 
of this system is characterized by a single invariant. Later, the octupoles are replaced by special nonlinear magnets, making the optics (again, theoretically) fully integrable (two invariants). Good agreement between the codes provides confirmation that the lattice element sequence is both read in and interpreted in a consistent manner and that basic single particle motion is modeled correctly by pyORBIT.
\subsection{Symplecticity Tests}  \label{sec: symplec_noSC}
In Hamiltonian mechanics, a transformation of the coordinates is said to be canonical if it preserves the form of
Hamilton's equations. A related result is that  dynamical evolution is itself a succession of infinitesimal canonical
transformations and therefore, the dynamical evolution from initial to final phase space coordinates is also a canonical
transformation. It can shown that a canonical transformation must satisfy a local analytic condition, the so-called
symplectic condition. Conversely, a transformation that satisfies the symplectic condition is canonical.     
The symplectic condition can be stated as follows:   
\begin{equation}
	\mtx{J}^T \mtx{S} \mtx{J} = \mtx{S}
	\label{symplectic_condition}
\end{equation}
 where $J$ is the Jacobian matrix of the transformation. Assuming the transformation is defined by the
 dynamical evolution map ${\cal M}$ connecting  the initial and final phase space coordinates
 ${\bf X}(s_0)$ and ${\bf X}(s_f)$  one has
\beq
    {\bf X}(s_f) =  {\cal M}{\bf X}(s_0)
 \eeq
\begin{equation}
	J_{k\ell} = \frac{\partial X_k(s_f)}{\partial X_\ell(s_0)} 
\end{equation}
where  ${\bf X}(s_i)$ represents the initial phase space vector of canonical coordinates
while $X_k(s_i)$ denotes individual  components of this vector. 
The symplectic matrix $\mtx{S}$ is defined as 
\begin{equation}
\mtx{S} = \left( 
\begin{array}{cccc}
\mtx{s} & 0 & \cdots & 0\\
0 & \mtx{s} & \cdots & 0\\
\vdots &  \vdots & \ddots &  \vdots \\
0 & 0 & \cdots & \mtx{s} 
\end{array}
\right) \;\; \mtx{s} = \left(
\begin{array}{cc}
  0 & 1 \\ 
 -1 & 0  
\end{array}
\right)
\end{equation}
Geometrically, the determinant of the Jacobian matrix  represents the ratio between final and initial differential (oriented)
phase space volumes enclosing corresponding points in phase space. Taking the determinant on both sides of
\ref{symplectic_condition}, and using $\det(S) = 1$, one finds $\det \,\mtx{J} = \pm 1$. However, in the limit of a
vanishingly small step along $s$ the dynamical transformation must converge to the identity $I$ i.e.
$\lim_{s\to 0} \mtx{J} = \mtx{I}$, and one concludes that $\det \mtx{J} = 1$ i.e. the phase space volume is a dynamical invariant.
This is the well-known Liouville's theorem.
 
A matrix is called symplectic if it satisfies Eq. (\ref{symplectic_condition}), known as the symplectic condition. While a unimodular Jacobian determinant is a necessary condition for symplecticity, it is not a sufficient test  in phase space of dimensions higher than two. 
We apply both  the determinant test and the full symplectic condition test Eq. (\ref{symplectic_condition}) as measures of the deviation from symplecticity. 
    
A residual deviation from exact symplecticity of the one-turn map computed by the code due to numerical round-off errors or approximations will translate into a long term violation of Liouville's theorem. Therefore, for accurate long term simulation of the dynamics, the Jacobian determinant of the one-turn maps generated by a tracking code should be as close as possible to 1. The finite size representation of floating point numbers sets a minimum for the achievable deviation.
Assuming 64-bit floating point representation (53-bit significance) the minimal deviation due to round-off is
$$ |\det(\mtx{J})-1| \simeq 10^{-14}$$

By tracking two or more test particles originating from closely neighboring points in phase space, one can compute partial derivatives using finite differences. Since the latter are approximations, deviations from the exact values of the partial derivatives will translate into larger apparent deviation from symplecticity. A rough error bound for the accuracy of the
derivative may be obtained from Taylor's theorem. For simple one-sided ( either forward or backward) differences 
 \begin{equation}
 \frac{\Delta X}{h} =  \frac{ f(X+h) -f(h) }{h}= f' + { \cal O}(h) 
 \end{equation}
 while for centered differences,
 \begin{equation}
 	\frac{\Delta f}{h} = 	\frac{f(X+h) -f(X-h)}{2h} = f' + { \cal O}(h^2) 
 \end{equation}
 Assuming that the Jacobian matrix is diagonally dominant with diagonal entries near unity and off diagonal entries of
 magnitude $<< 1$ it can be shown that  
 \begin{equation}
 	1 - h < \det (\mtx{J} + \mtx{h}) < (1 + 2h + n h^2)^{n/2}  
  \end{equation}
 where $n$ is the matrix dimension and it is assumed that the magnitude of the error affecting all the Jacobian entries is
 about the same.  On the basis of this bound, using centered differences and $h \sim 10^{-3}$, deviations from unity of
 the determinant on the order of $10^{-6}$ are to be expected; larger deviations would be indicative either of code
 issues (i.g. incorrectly computed map) or of the presence of chaotic motion.   
 
 The previous discussion has implicitly assumed well-behaved, smooth differentiable dynamical maps. It should be
 obvious that to the extent that maps associated with chaotic motion are not (practically) differentiable, the Jacobian
 determinant test is expected to fail in chaotic regions. In a typical accelerator the dynamics is smooth in the vicinity of the
 reference orbit. Chaotic motion is observed as one moves away from reference trajectory, where nonlinearity become
 increasingly significant. An abrupt deviation of the Jacobian determinant away from unity may be used to delineate the
 dynamic aperture boundary.

 We now discuss the results of the symplectic tests of single particle tracking with pyorbit. Since this report is limited to
 4D (transverse) space charge studies, we consider 4D symplectic tests both here and in a later Section where space
 charge is turned on\ref{sec: sympl_SC}. 
 To perform numerical tests of symplecticity, it is best to work in normalized Floquet canonical coordinates. The latter are defined as follows     

\beq
\mathbf{X} =  \left[ \fr{x}{\sqrt{\bt_x}},\fr{(\bt_x x' + \al_x x)}{\sqrt{\bt_x}},  \fr{y}{\sqrt{\bt_y}},
  \fr{(\bt_y y' + \al_y y)}{\sqrt{\bt_y}} \right]   \label{eq: Floquet}
\eeq
In these coordinates, all components of ${\bf X}$ have dimension $[\sqrt{L}]$ and the elements of the Jacobian
matrix $\mtx{J}$ are  dimensionless.

 We first present results of the symplecticity tests for the linear lattice. Figure \ref{fig: sympl_llnear} shows the deviation from unity of
 $\det \mtx{J}$ and the maximum norm $|\mtx{J}^T \mtx{S} \mtx{J} - \mtx{S}|$ as functions of the transverse amplitude for this linear lattice.

In both figures, the deviations are shown for three small increments (0.01, 0.001, 0.0001) of the initial amplitude. 
\bfig
\centering
\includegraphics[scale=0.42]{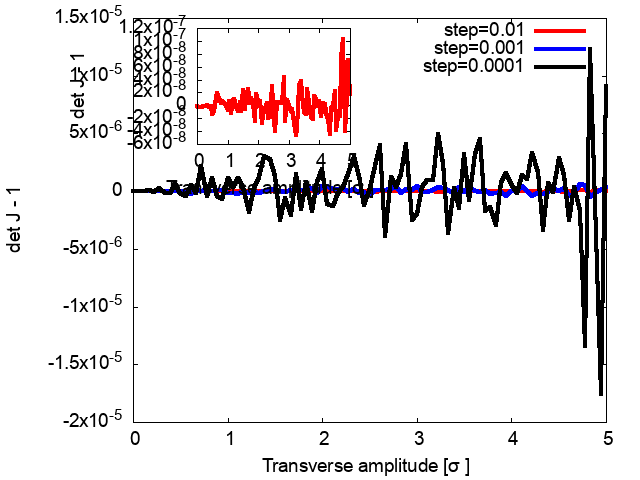}
\includegraphics[scale=0.42]{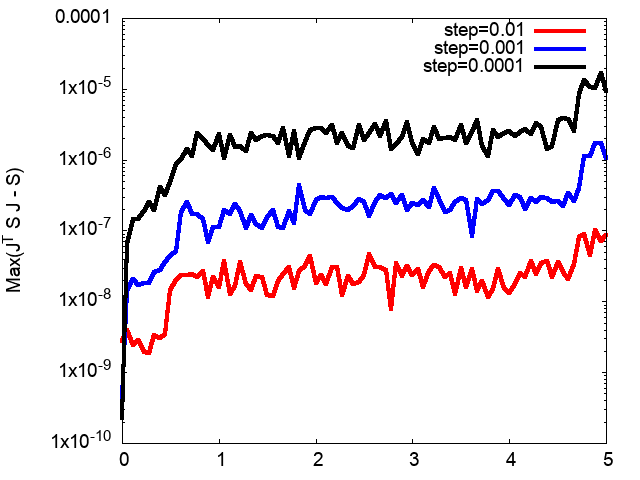}
\includegraphics[scale=0.42]{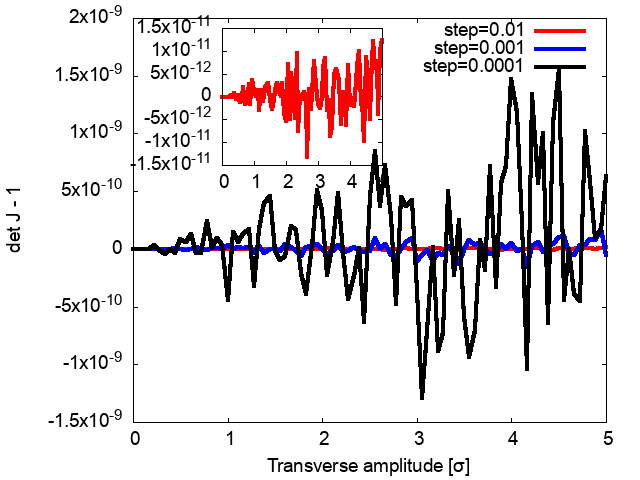}
\includegraphics[scale=0.42]{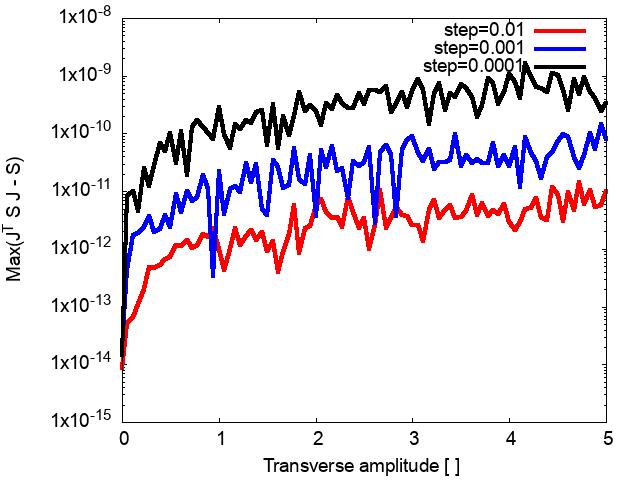}
\caption{Linear lattice. Difference $|\det \mtx{J}-1| $ (left) and maximum norm of
    the matrix $(\mtx{J}^T \mtx{S} \mtx{J} - \mtx{S})$ (right) as functions of the transverse amplitude in units of
    the rms beam size. Top: MADX, Bottom: pyORBIT. The left plot in both rows  has insets showing the smallest values
of the difference which result for the step size 0.01. }
\label{fig: sympl_llnear}
\efig
We find that the deviation from unity $\det \mtx{J} - 1$ is nearly $10^{-7}$ with MADX. For pyORBIT the deviation is at most 10$^{-11}$ for the at the largest step size of $0.01$ and increases for smaller step sizes. Similarly, the maximum
norm  $ | \mtx{J}^T \mtx{S} \mtx{J} - \mtx{S}|$ is $\sim 10^{-8}$)  with MADX and ($\sim 10^{-12}$) with
 pyORBIT for the step size $0.01$ and increases by an order of magnitude with each successive decrease in step size.o The larger deviation from unity of the Jacobian determinant observed with MADX appears to be related to its handling
 of dipole edge focusing, which is turned off in pyorbit.

Results for the lattice with the octupole string in IOTA are shown in Figure \ref{fig: sympl_oct}.
Up to amplitudes of 4$\sg$, $|\det[ \mtx{J}] - 1|$ at step size $0.01$ is  $\sim 10^{-8}$ with MADX and $\sim 10^{-4}$
with pyORBIT. The deviation  increases steeply and is the largest at higher betatron amplitudes with both codes. For the
smaller step of 0.001, the largest deviation over
0-5 $\sg$ is $\sim 10^{-4}$ and can be considered the optimum step size; the deviations are larger again at the
smaller step size 0.0001. The same pattern is observed for the maximum norm
$|\mtx{J}^T \mtx{S} \mtx{J} - \mtx{S}|$.

\bfig
\centering
\includegraphics[scale=0.42]{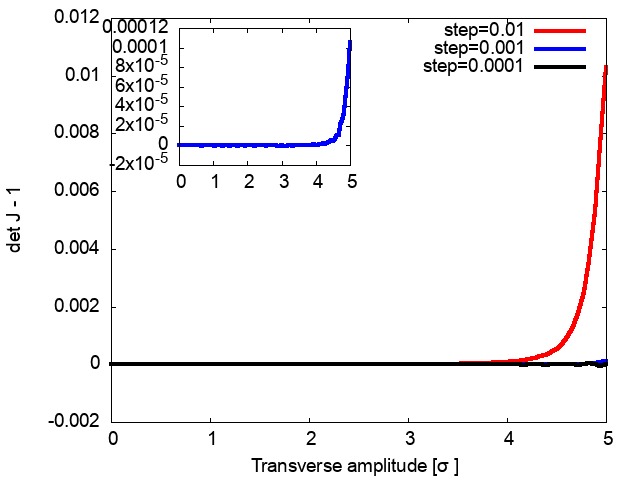}
\includegraphics[scale=0.42]{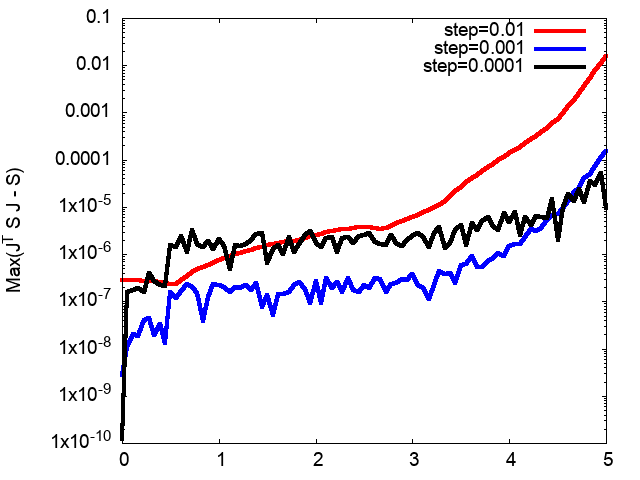}
\includegraphics[scale=0.42]{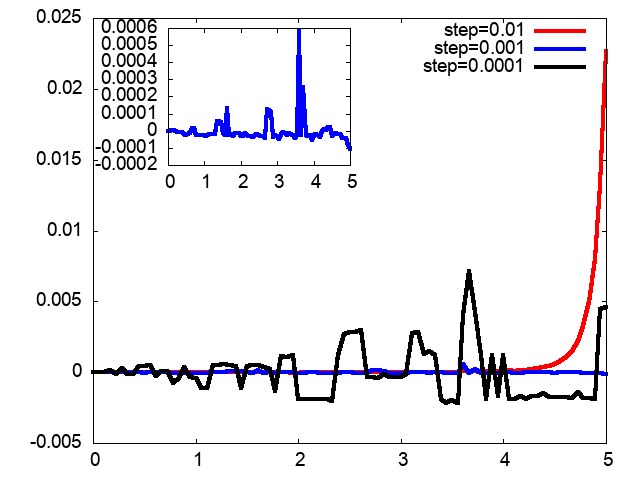}
\includegraphics[scale=0.42]{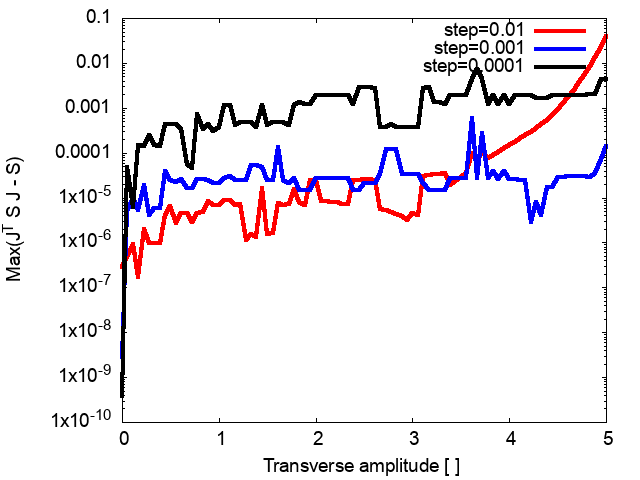}
\caption{Lattice with octupoles. $(\det \mtx{J} -1)$ (left) and maximum element of
    the matrix $ \mtx{J}^T \mtx{S} \mtx{J} - \mtx{S} $ (right) as functions of the transverse amplitude in units of
    the rms beam size. Top: MADX, Bottom: pyORBIT}
\label{fig: sympl_oct}
\efig
\subsection{Tune Footprint}
A meaningful way to validate pyORBIT against MADX is to compare tune footprints. To do so, particles are initialized
 uniformly from $(0, 0, 0, 0)$ to $(5\sigma_x, 0, 5\sigma_y, 0)$ in the $x$ and $y$ planes. They are then  tracked for
5000 turns and the transverse positions $(x, y)$ of every particle are  recorded for each turn at a fixed location. For each
particle, the fractional part of the tunes $Q_x$ and $Q_y$ are obtained by performing a Fast Fourier Transform
(FFT) of $x(n)$ and $y(n)$.  $Q_x$ and $Q_y$ are presented as a scatter plot in tune space.
\begin{figure}%%% [h]
  \centering
  \includegraphics[scale = 0.20]{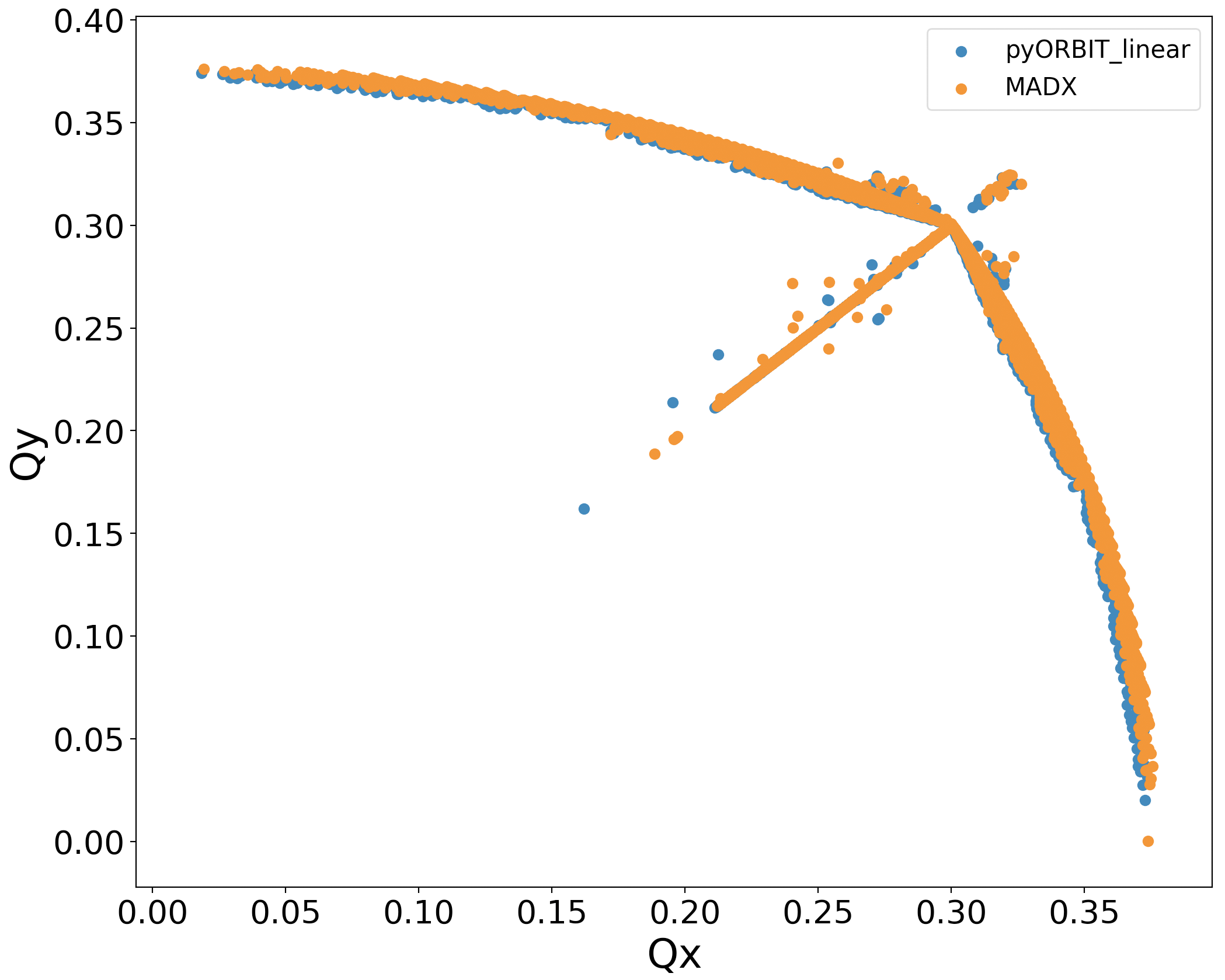}
  \includegraphics[scale = 0.2]{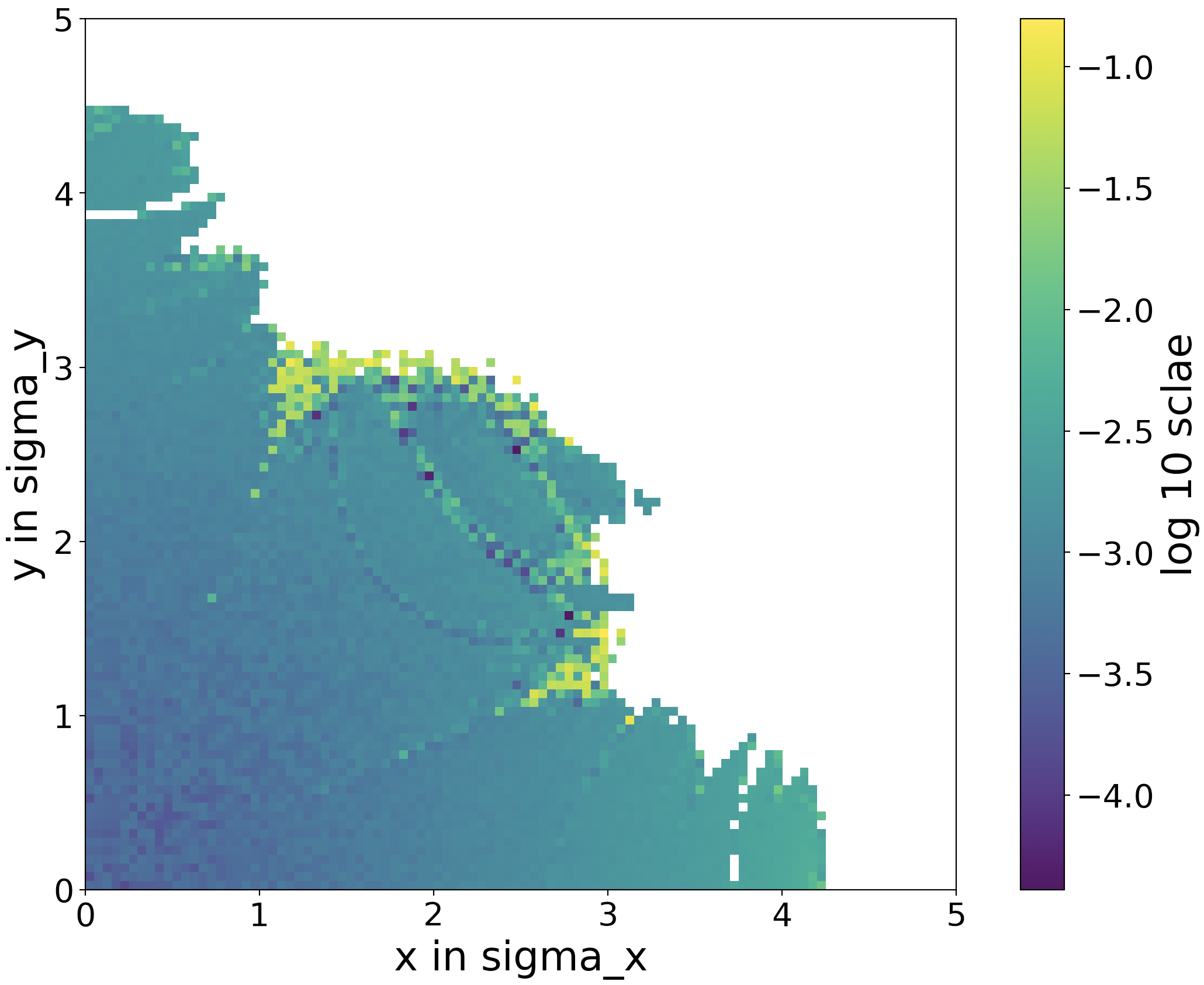}
  \caption{Left:  Tune footprint of the IOTA lattice with octupoles, obtained with pyORBIT and MADX.
  Right: Difference in tune shifts (on a log scale) between pyORBIT and MADX as functions of the initial positions
of the test particles.}
\label{fig:tune_fft}
\end{figure}
The left plot in Figure \ref{fig:tune_fft} shows the tune footprint with  octupoles added to the IOTA lattice.
 Both MADX and pyORBIT show a similar tune spread, which serves as a cross-check of the pyORBIT single particle tracking model. To make a more meaningful quantitative comparison between the codes, we present the difference in the tune shift computed by the two codes, and  it is computed as
\beq
dQ = \mid Q_{x, P} - Q_{x, M} \mid + \mid Q_{y, P} - Q_{y, M} \mid
\eeq
where e.g. $Q_{x, M}, Q_{x, P}$ are the horizontal  tunes calculated with MADX and pyORBIT respectively. 
The right plot in Figure \ref{fig:tune_fft} shows the tune difference $dQ$ for all test particles as functions of their initial positions.
 Figure \ref{fig:tune_fft}b shows the tune difference $dQ$ for all test particles as function of their initial positions.
 For most test particles the difference between the tunes predicted by MADX and pyORBIT falls within the range
 $10^{-4}: 10^{-3}$, which is comparable to the resolution of the FFT sampling. The result provides another confirmation of the correctness of single particle tracking in pyORBIT. 
 Particles exhibiting the largest discrepancies are located close to the dynamic aperture boundary, as shown in
Fig. \ref{fig:tune_fft}. Beyond this boundary, the particles are lost and  $\log dQ$ is undefined. All particles in the tune
the  footprint are initialized at coordinates $(x, 0, y, 0, 0, 0)$. The boundary defined by large deviations seen in Fig. \ref{fig:tune_fft}  corresponds to the 4D dynamic aperture in Figure \ref{fig:DA_oct}a. Small 
 differences between the tune calculated by MADX and pyORBIT are likely due to increasingly unstable motion in the
 vicinity of the dynamic aperture boundary. We remark in passing that the fractal nature of this boundary is made
 evident by the tune difference diagram and that a similar procedure is used in a frequency map analysis except that in
 that case the same code is used to calculate tune differences between neighboring particles. 
\subsection{Dynamic Aperture Tests}
The 4D, 5D and 6D dynamic apertures are calculated for the IOTA lattice with and without octupoles using both MADX
and pyORBIT. Circular physical apertures with radius $25$ mm are assigned to all elements. $5000$ particles are
initialized using one of the following procedures: 
 The 4D dynamic aperture is calculated with the rf cavity turned off and particles are initialized at 
 ($x_i$, 0, $y_i$, 0, 0, 0) so that the motion is restricted to transverse 4D phase space. 
 The 5D dynamic aperture is calculated with the rf cavity turned off, and particles are initialized similarly but with a
constant  momentum offset of 1 $\sg_p$ so that  chromaticity can  affect the dynamic  aperture.
The rf cavity is turned on for the 6D  calculations with the same initial coordinates so that the effects of 
 synchrotron motion are included. 

A perfect lattice is used;  no errors, e.g. field or alignment errors, are included. 
Particles are tracked for 10000 turns, and the initial coordinates in the $x$-$y$ plane of all surviving particles are
recorded. The largest excursions of the surviving particles yield an upper bound  of the dynamic
aperture.
\begin{figure}%%% [h]
  \centering
  \includegraphics[scale = 0.22]{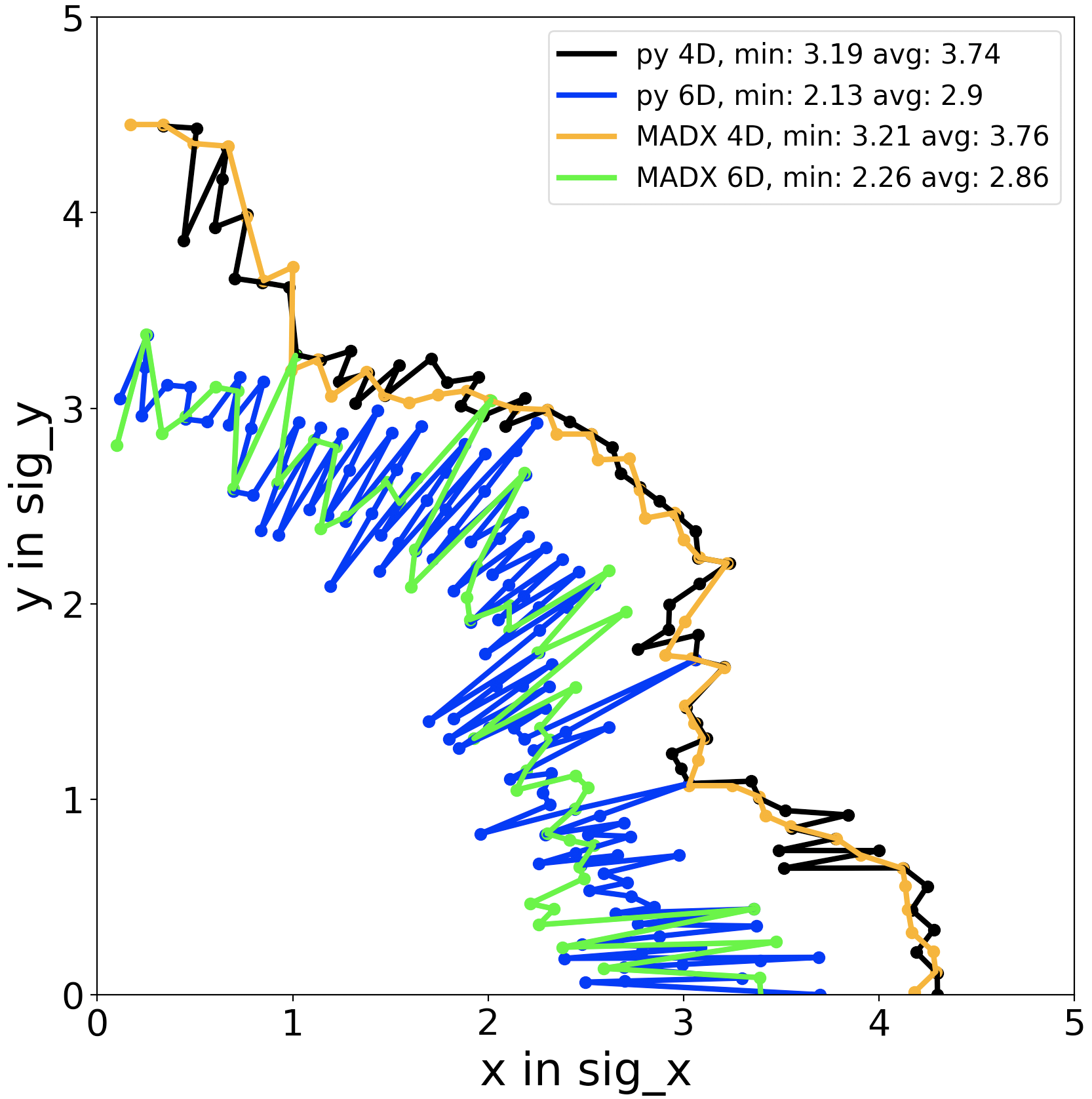}
  \includegraphics[scale = 0.220]{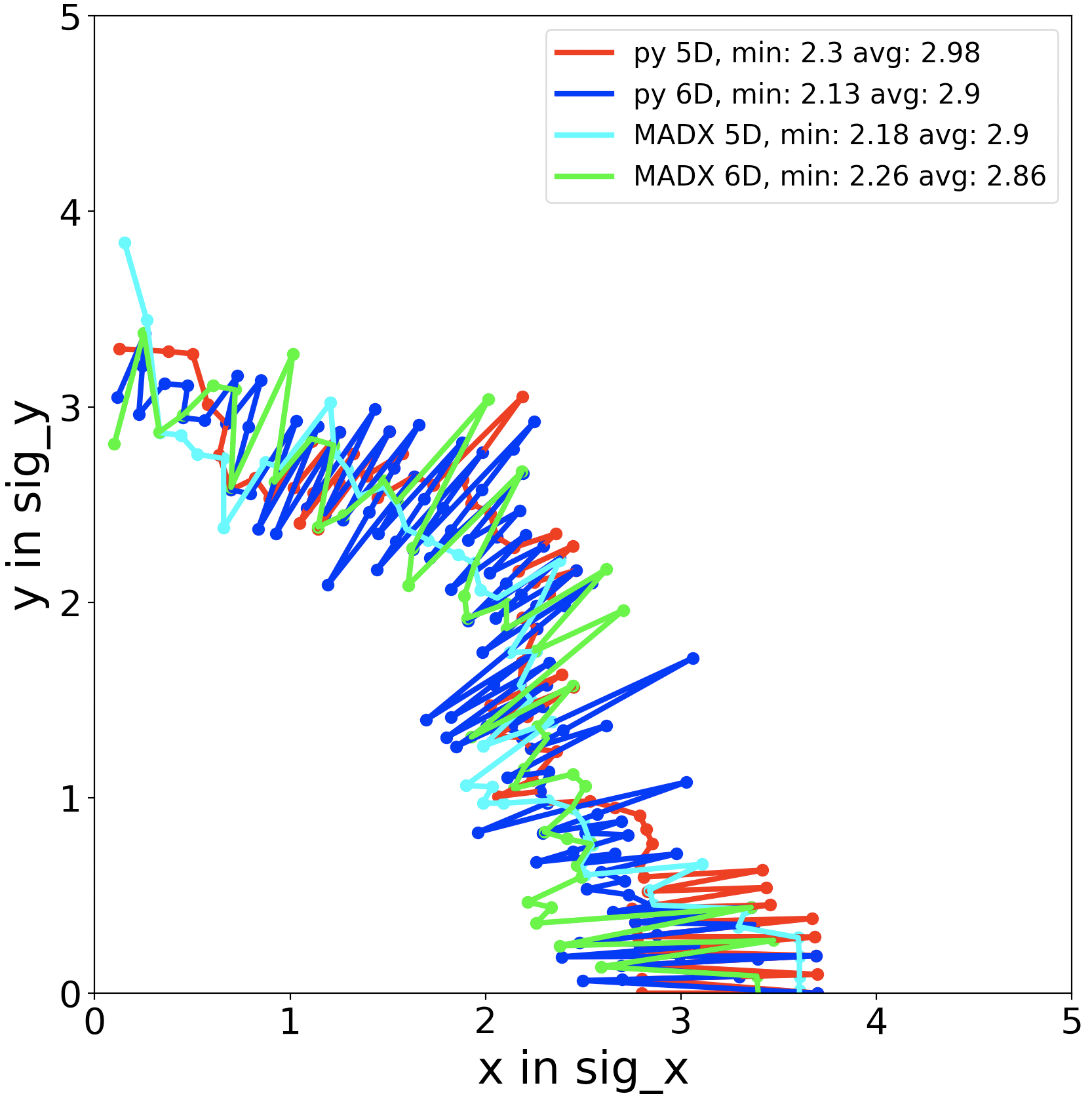}
  \caption{4D, 5D, and 6D dynamic apertures (expressed in units of the rms beam sizes)
    for IOTA with octupoles from pyORBIT and MADX.
    Left:  Comparison of 4D and 6D apertures, Right:  Comparison of 5D and 6D apertures, they are
    nearly the same. }
\label{fig:DA_oct}
\end{figure}
In the purely linear lattice,  the horizontal dynamic aperture is approximately $5\sigma_x$ and
 $6\sigma_y$ vertically.  As expected, the dynamic aperture is about the same as the physical aperture at the peak beta locations $\hat{\beta}_{x,y}$. Results from pyORBIT and MADX are  in good agreement.

 Results of dynamic aperture calculations with octupoles added to the linear lattice are shown in
Figure \ref{fig:DA_oct} and summarized in Table \ref{table: DA_oct} of minimum and average aperture sizes expressed
in terms of rms beam sizes.
\begin{table}
\begin{center}
\begin{tabular}{ |c|c|c||c|c| } 
 \hline
 Aperture type & \multicolumn{2}{|c||}{ Minimum DA}   &  \multicolumn{2}{|c|}{Average DA} \\ 
 &  {\sc pyORBIT} &    {\sc MADX} & {\sc pyORBIT} &    {\sc MADX} \\ \hline
 $4$D &  3.19 & 3.21 &  3.74  &  3.76  \\ 
 $5$D &  2.3 & 2.18 & 2.98   &  2.9  \\ 
 $6$ D&  2.13 & 2.26  & 2.9  & 2.86  \\
 \hline
\end{tabular}
\caption{Minimum and average dynamic apertures with octupoles calculated using pyORBIT  and MADX.}
\label{table: DA_oct}
\end{center}
\end{table}

Referring to Table \ref{table: DA_oct} and to Figure \ref{fig:DA_oct}, there is good agreement between MADX and
pyORBIT, providing additional confirmation of the soundness of the single particle tracking model in pyORBIT.
The dynamic apertures for 6D and 5D tracking are generally much smaller than for 4D. For an emittance of
0.3$\mu m$
the latter is about 3 $\sigma$. This indicates that with a transverse Gaussian distribution, truncation at 3$\sigma$
should prevent particle loss. However, significant particle losses remain likely in the presence of space charge and for a more realistic lattice that includes alignment errors. 

\subsection{Hamiltonian Test}

In the original paper on the integrable lattices \cite{Danilov} that provided the motivation for IOTA, the derivation assumes that in the absence of nonlinearity the $\beta$-function of the ring is symmetric i.e. $\beta_x(s) = \beta_y(s) = \beta(s)$ within the nonlinear insertion region. In principle this can be realized at low energy with radial transverse focusing provided by solenoids as this makes the beta function symmetric everywhere. However, as pointed out in the paper, the $\beta$-function symmetry requirement applies only within the nonlinear insertion region. With the nonlinearity confined to region $(s_1,s_2)$ the conditions $\beta_x(s_1) = \beta_y(s_1) = \beta_x(s_2) = \beta_y(s_2))$ are sufficient to ensure this outcome.     
In IOTA, focusing in the linear portion of the ring is provided by quadrupoles. The optics is designed such that the map corresponding to that section is that of a thin axially symmetric lens. For purposes of analysis it is easier to assume ring-wide radial focusing; however to the extent that the $\beta$-function is symmetric within insertion region, the dynamics of the two situations is equivalent. 

Expressed in terms of the independent variable $s$, 
the ring Hamiltonian has the analytical form \cite{Danilov} 
\beq
H = \half(p_{x}^2 + p_{y}^2)  + \frac{k(s)}{2}(x^2 + y^2)  +  V(x, y, s)
\eeq
where 
\begin{equation}
	k(s) = 
\begin{cases}
0  &   s_1 < s < s_2 \\ 
k  &  \text{elsewhere} 
\end{cases}
\end{equation}
is the linear radial focusing and $V(x,y,s)$ is a potential. The latter satisfies 
\begin{equation}
	V(x,y, s) = 
	\begin{cases}
		V(x,y,s)  &   s_1 < s < s_2 \\ 
		0  &  \text{elsewhere} 
	\end{cases}
\end{equation}
i.e. it vanishes outside of the nonlinear insert. An octupole potential can be written as
\beq
V(s) = \fr{K_3(s)}{4!} [  x^4 +  y^4  -  6 x^2 y^2 ], \;\;\; K_3 = \fr{1}{(B\rho)} \fr{\del^3 B_y}{\del x^3}    \label{eq: V_oct}
\eeq
where the strength $K_3(s)$ (which has the definition as in MADX) varies with $s$.
If the strength of each octupole in the string is chosen to vary  as the inverse cube of the beta function 
in the section $s_1 < s < s_2$ where 
$\bt_x(s) = \bt_y(s) = \bt(s)$  \cite{Danilov}, then
\beq
K_3(s) = \kappa / \bt(s)^3    \label{eq: kappa}
\eeq
where $\kappa$ is a constant independent of $s$. 
 Expressed in terms of Floquet coordinates  $x_N= = x /\sqrt{\bt_x}, y_N= = y /\sqrt{\bt_y}$, and
transforming to the phase advance $\psi = \int \; ds/\bt(s)$  as the independent variable,
the rescaled potential is
\beq
U = \bt(s) V((s) =   \fr{1}{24} \kappa[ x_N^4  +  y_N^4 - 6 x_N^2 y_N^2 ]
\eeq
which is independent of the variable $\psi$.

Strictly speaking a purely transverse potential with a smooth longitudinal variation is not physically realizable; however, it can be reasonably  well-approximated by a string of octupole magnets of different apertures; for that purpose, the IOTA octupole insert design uses 17 equally spaced magnets.  

Figure \ref{fig: betaxy} shows the $\beta$-functions in the IOTA ring; the octupoles occupy the 1.7 m section
from $s=33$ m to $s=35$ m.  The use of a sequence of 
discrete magnets is expected to result in small fluctuations of the idealized invariant Hamiltonian. Lattice imperfections in the linear part of the ring causing violations of the condition $\beta_x = \beta_y$ within the nonlinear insert region will have a similar effect.
  \bfig
  \centering
  \includegraphics[scale=0.5]{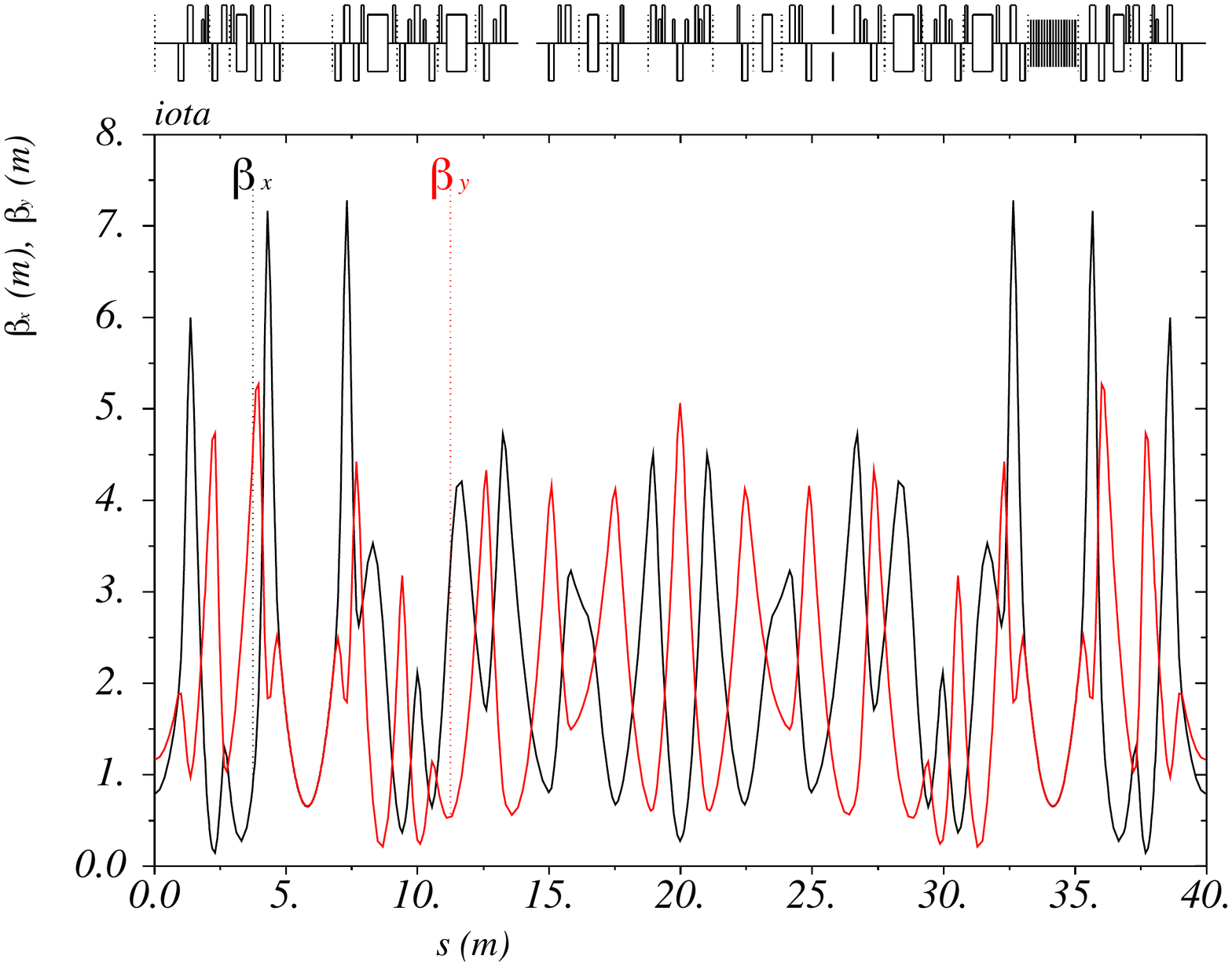}
  \caption{Beta functions in the IOTA ring; the octupoles are in the region from 33.3 to 35m.}
  \label{fig: betaxy}
  \efig

  \bfig
  \centering
  \includegraphics[scale=0.35]{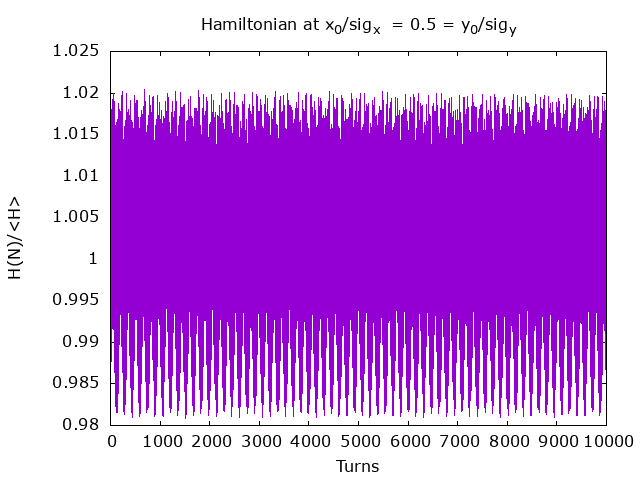}
  \includegraphics[scale=0.45]{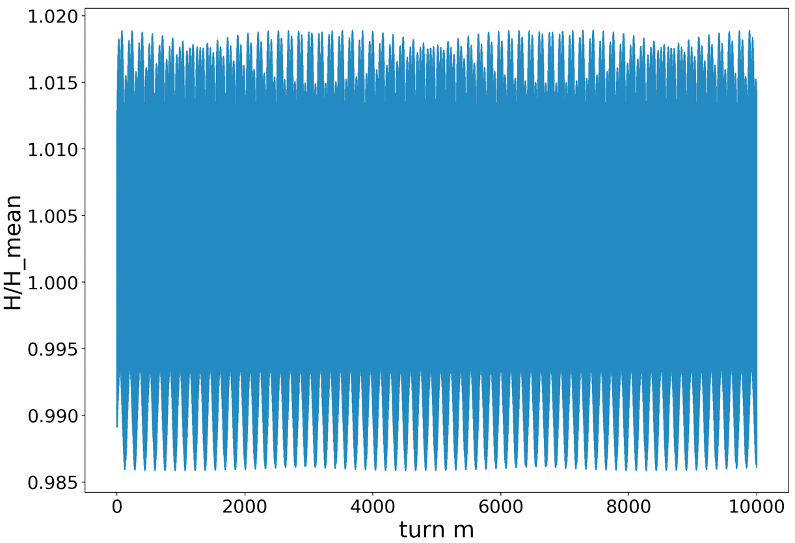}
  \caption{Variation in the single particle Hamiltonian relative to its initial value in MADX (left) and pyORBIT
    (right).    Initial coordinates are $x/\sg_x = 0.5 = y/\sg_y$, $x' = 0 = y'$.}
  \label{fig: Ham_var}
  \efig
  Figure \ref{fig: Ham_var} shows that the relative variation of the Hamiltonian invariant for a small amplitude particle predicted by both MADX and pyORBIT  is approximately $\pm 2\%$. This variation increases for larger amplitudes and both codes remain in agreement. 
%
%-----------------------------------  
\section{pyORBIT Space Charge Model}  
%-----------------------------------
{\sc pyORBIT} is a particle-in-cell code in which macro-particles represent charged particles in a bunch. The
macroparticles are deposited on a grid and a smoothed density is extracted by  interpolation between the grid points.
The electric potential is found by numerically solving Poisson's equation in the beam rest frame which thereby allows
evaluation of the space charge forces on the macro-particles. The details of the physical (as opposed to numerical) space
charge force depend on the bunch intensity, the particle distribution and boundary conditions. In this report we
consider only the transverse space charge forces and we assume open boundary conditions. The transverse
distributions are assumed to be Gaussian since that is experimentally well-founded. The transverse forces vary along the
bunch length due to the change in local density, so the longitudinal distribution also matters.

\begin{figure}%%% [h]
  \centering
\includegraphics[scale = 0.14]{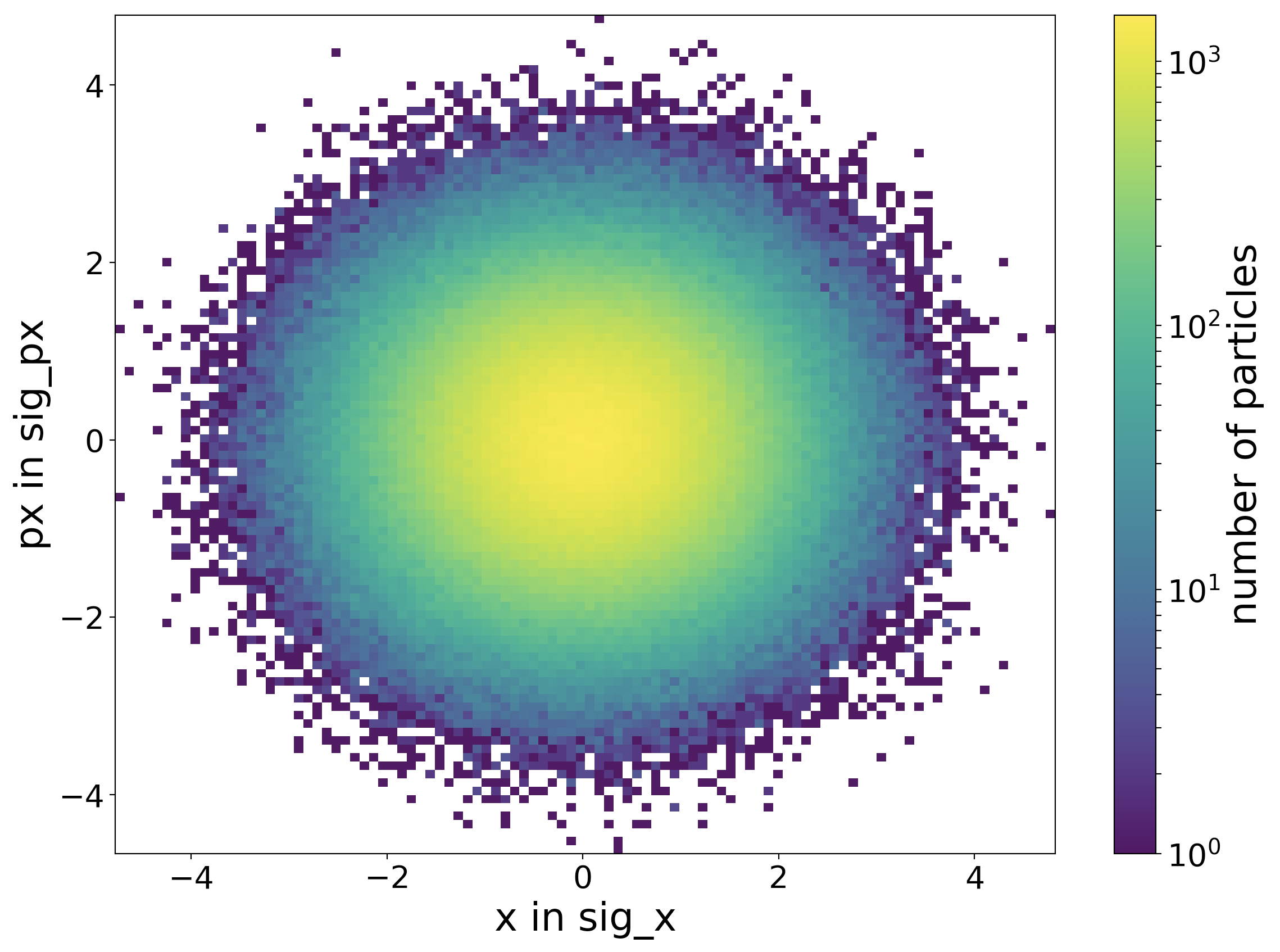}
  \includegraphics[scale = 0.14]{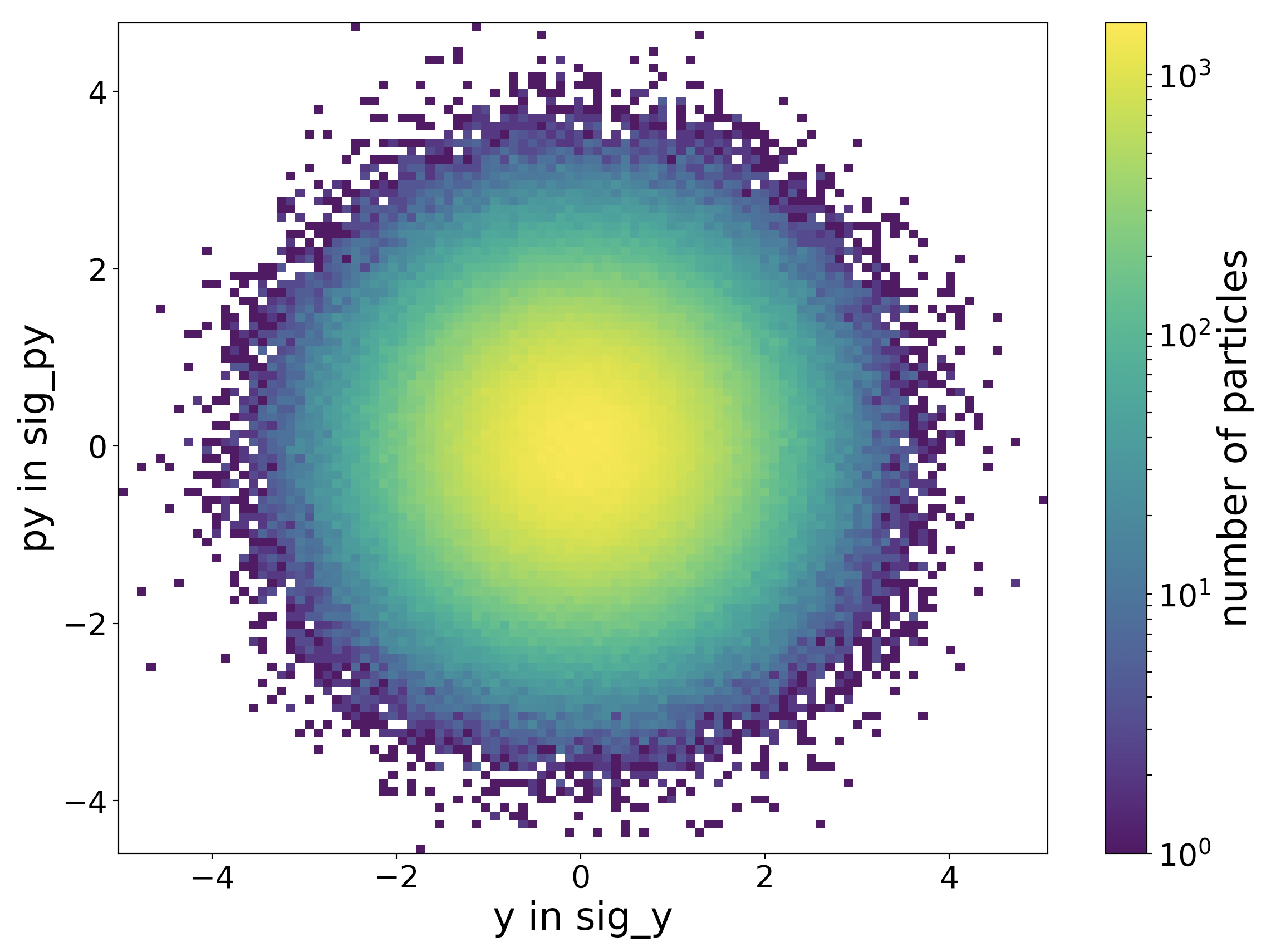}
  \caption{Particle distribution in x-pz (left)  and y-py plane (right) }
  \includegraphics[scale = 0.14]{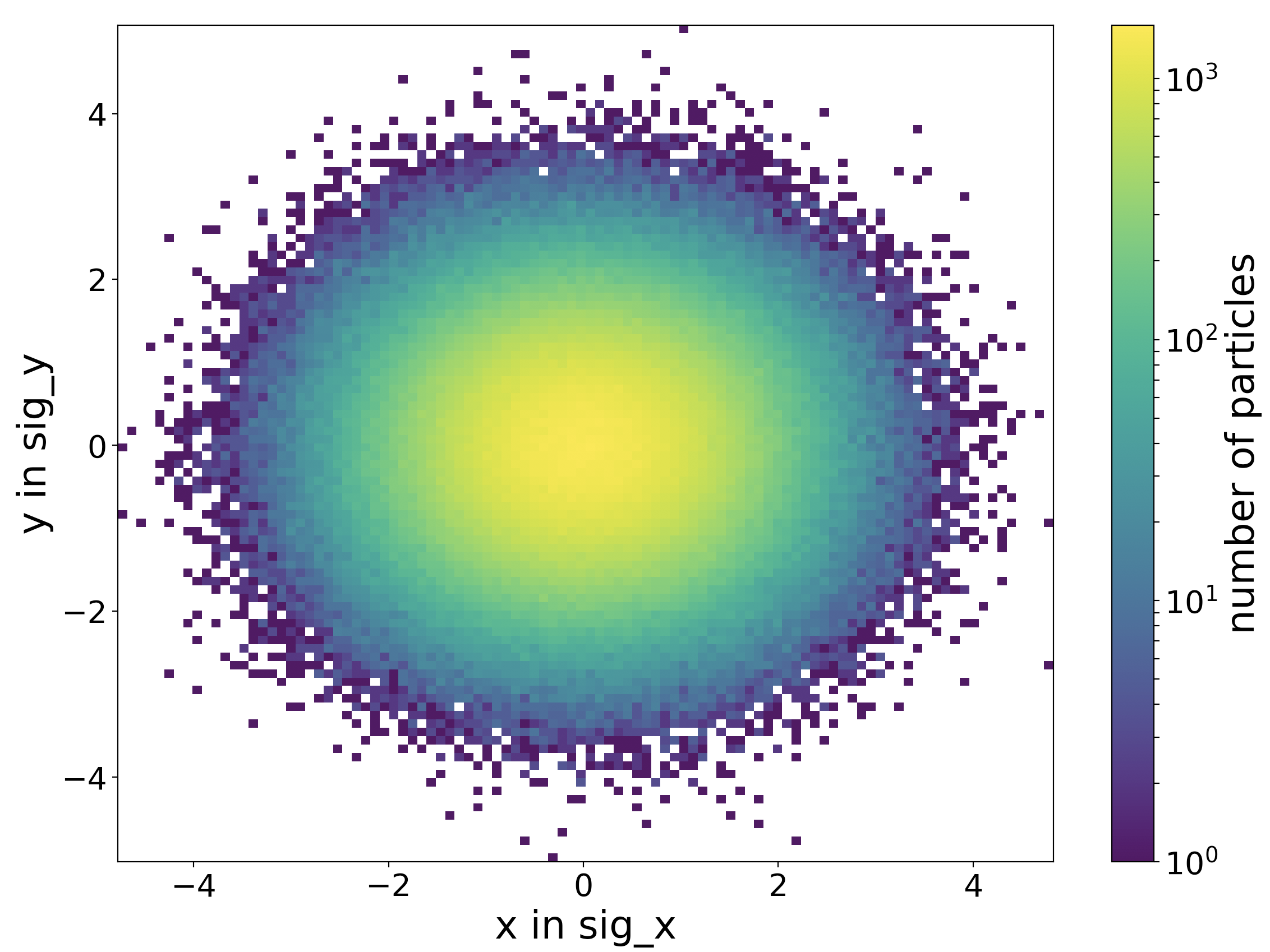}
  \includegraphics[scale = 0.14]{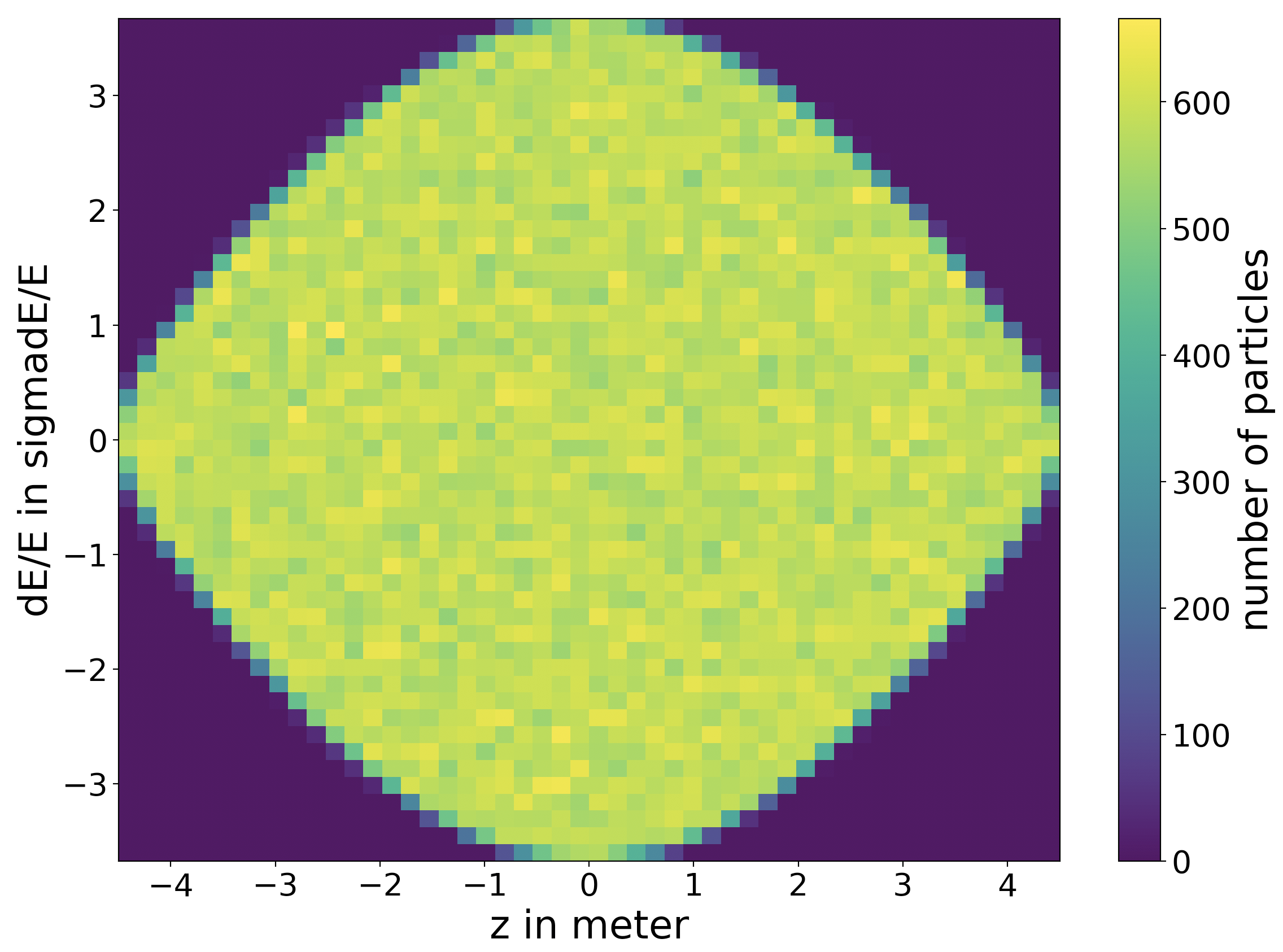}
  \caption{Particle distribution in x-y (eft) and z-dE plane (right)}
\label{fig:distribution_example}
\end{figure}

In our simulations, a waterbag distribution is used in the longitudinal  plane. To prevent particles leaking out of the
bucket when longitudinal space charge forces (to be reported in a separate study) are considered, particles are
uniformly distributed within an inner Hamiltonian contour inside the separatrix defined by the RF cavity. An example
using $10^6$ macro-particles is plotted in Figure \ref{fig:distribution_example}.

\subsection{Symplectic tests with space charge}
\label{sec: sympl_SC}

\bfig
  \centering
  \includegraphics[scale = 0.42]{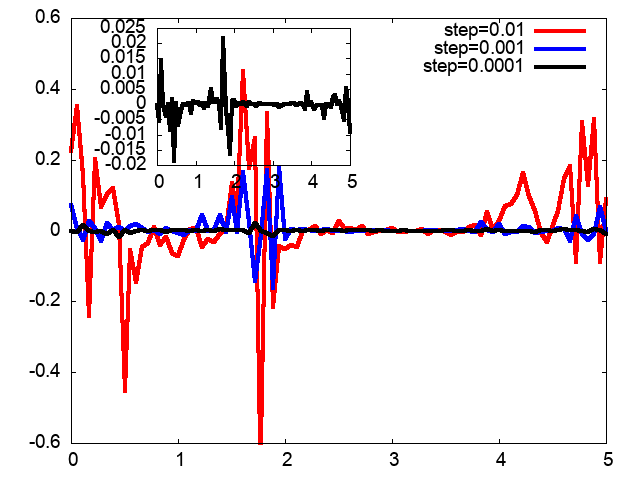}
  \includegraphics[scale = 0.42]{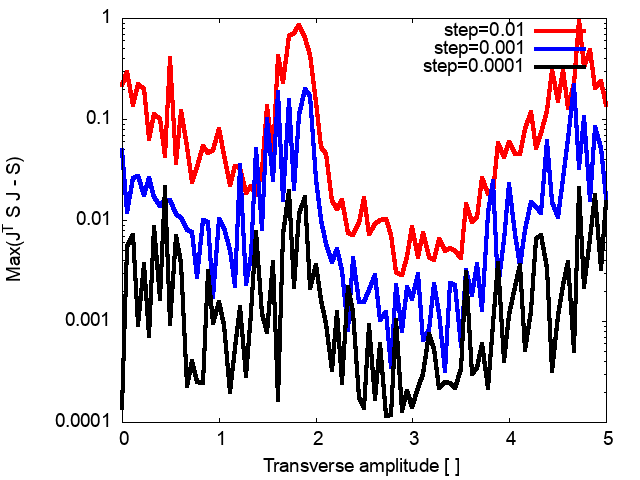}
  \caption{Symplecticity tests with space charge. Difference ${\mathrm{det}[\bf J}$] - 1 (left) and maximum norm of
    the matrix $ {\bf J}^T {\bf S} {\bf J} - {\bf S}$ (right) as functions of the transverse amplitude in units of
    the rms beam size.}
  \label{fig: sympl_SC}
  \efig
  Figure \ref{fig: sympl_SC} shows the results of symplecticity tests with space charge. The largest deviation 
 $\det{\bf J}-1$ is $0.4$ using a step size of $0.01$, it falls to 0.15 at step size $0.001$ and is an order of magnitude
 smaller (0.02) at step size $0.0001$. The maximum norm  $|{\bf J}^T {\bf S} {\bf J} - {\bf S}|$ exhibits a similar behavior and reaches its smallest value at step size $0.0001$.
 Compared to the cases with octupoles but without space charge in Figure \ref{fig: sympl_oct}, the deviations from symplecticity are at least two orders of magnitude larger. Another difference is that the largest deviations occur
 at a lower betatron amplitude (about 2$\sg$) and do not increase with amplitude. 

The fact that the deviations from symplecticity in pyORBIT in the presence of space charge are significant is not
surprising as it is observed in most PIC codes to date. This state of affairs makes these codes not suitable for long
term tracking.

\section{Emittance Growth and Particle Loss}
\label{sec: emit_loss}  

Predictions of emittance growth and particle loss are two of the many reasons for using a space charge code.
We first discuss the results of emittance growth without space charge to estimate the background growth and then
at low and full bunch intensity to observe the impact of increasing space charge effects.
The top row in Fig. \ref{fig: emit_No_Lo_Hi_SC} shows the emittance in all three planes without space charge and at a
low bunch charge of 10$^6$ protons, five orders of magnitude lower than the full intensity. There is no particle loss in
either case. 
\bfig
\centering
\includegraphics[scale = 0.40]{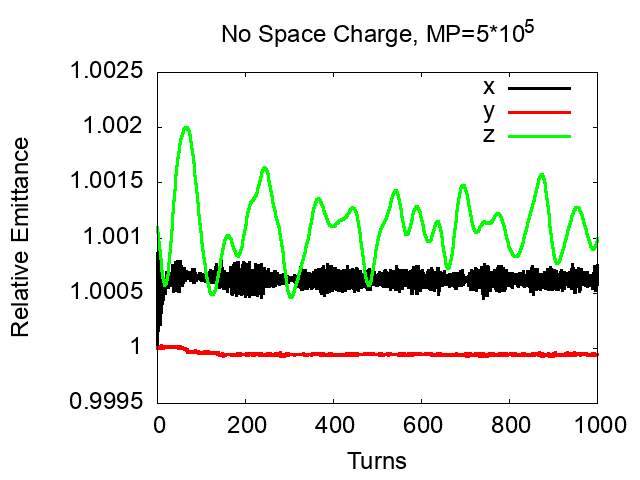}
\includegraphics[scale = 0.40]{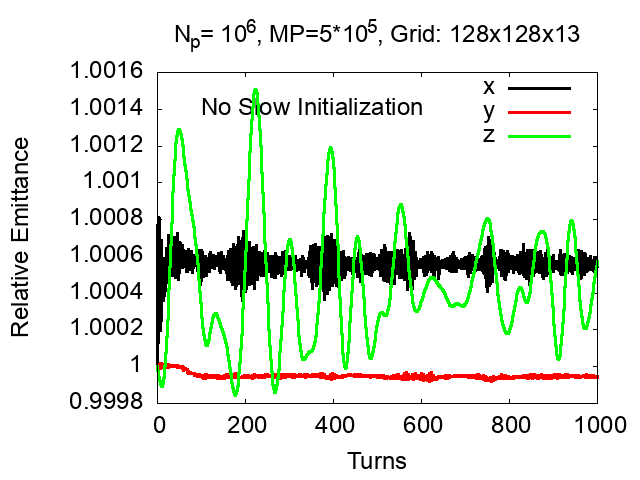}
\includegraphics[scale = 0.50]{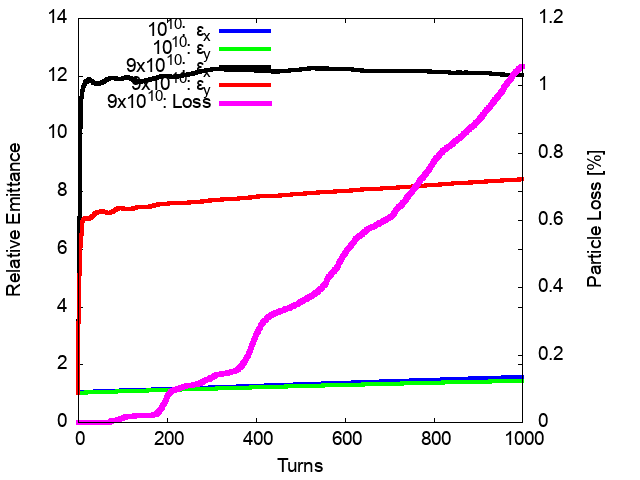}
\caption{Top:  Emittance growth in each plane without space charge (left) and at a low bunch intensity of
  $10^6$ (right). Bottom: Emittance growth (left vertical scale) at intensities 10$^{10}$ and $9\times 10^{10}$
and the particle loss (right vertical scale) at the higher intensity. There is no loss at 10$^{10}$ over this time
  scale, hence is not shown. }
\label{fig: emit_No_Lo_Hi_SC}
\efig
The influence of dispersion and momentum spread is visible in the transverse emittance. The oscillations in both the
horizontal and longitudinal emittance have the same period.  The background noise in the transverse emittances is
 less than 0.1 \%.  
 The bottom plot in this figure shows the emittance growth at bunch intensities 10$^{10}$ and $9\times 10^{10}$.
 At the lower
 intensity, the emittance grows by 20\% and there is no loss. At the higher intensity, the emittance growth is dramatic,
 up to 12 times ($\eps_x$) and 7 times ($\eps_y$). The growth occurs over the first few turns and quickly reaches a
 plateau. The predicted total loss over 1000 turns is $\sim 1\%$, which is unacceptably high for such a short time scale.

  \subsection{Initial rms matching}
  \label{sec: match}

  As seen in Fig. \ref{fig: emit_No_Lo_Hi_SC}, the  mismatch between the bare lattice functions and those with space
  charge   leads to a rapid initial emittance growth and particle loss at full intensity. 
  Properly matching a lattice to space charge induced changes requires matching the Twiss functions 
 so that the beam envelopes have the space charge equilibrium values  all around the ring.  
 One way of extracting the matched Twiss functions is to solve the envelope equations for the stationary  rms
 sizes and use the initial emittances to find these matched functions. 
 Instead here we track the unmatched macroparticle distribution
  for 1000 turns. At the end of this period, we extract the statistical Twiss functions
  e.g in the horizontal plane as  (here $\dl = \Dl p/p$)
  \beqr
  \lan  \bt_x \ran &  = &  \fr{\lan x^2 \ran}{\sqrt{ \lan x^2 \ran \lan (x')^2 \ran -  \lan (x x')^2 \ran}}, \;\;\;
  \lan \al_x  \ran = - \fr{ \lan x x' \ran}{\sqrt{ \lan x^2 \ran \lan (x')^2 \ran -  \lan (x x')^2 \ran}}  \\
\lan  D_x  \ran & = & \fr{\lan x \dl \ran}{\lan \dl^2 \ran}, \;\;\;\;  \lan D_x'  \ran = \fr{\lan x' \dl \ran}{\lan \dl^2 \ran}
  \eeqr
\bfig
\centering
\includegraphics[scale = 0.50]{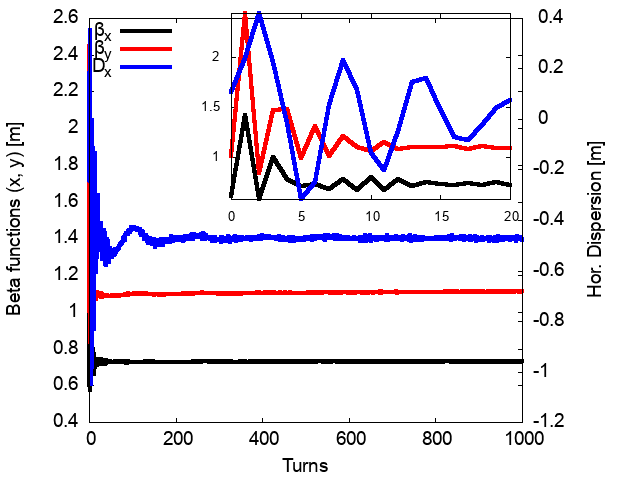}
\caption{The statistical beta functions (on the left vertical scale) and dispersion (right vertical scale) over time. 
  The inset shows that  the beta functions  relax to their equilibrium values within 20 turns, the dispersion relaxes over
about 200 turns.}
\label{fig: beta_orig}
\efig
Figure \ref{fig: beta_orig} shows the evolution of the beta functions. There are sharp transients as the beam adjusts
to the space charge forces but the beta functions reach stable values which turn out to be close to their initial values
within 20 turns; the same 
time  scale as for the emittance as seen in Fig. \ref{fig: emit_No_Lo_Hi_SC}. The initial beta functions are
$(\bt_x, \bt_y) = (0.79, 1.16)$m and the final values after 1000 turns are (0.73, 1.11)m. The dispersion function
relaxes over a longer time scale and changes more significantly from -0.24m to -0.43m.
The lattice is then re matched with these values at the initial point in the lattice and is used in the subsequent
tracking calculations discussed below. 
\bfig
\centering
\includegraphics[scale = 0.40]{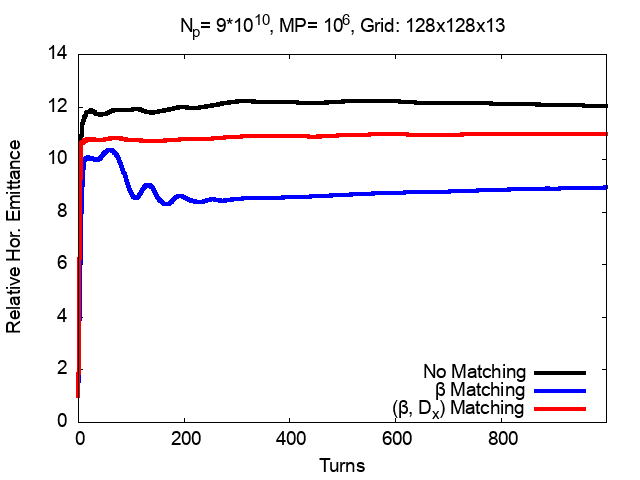}
\includegraphics[scale = 0.40]{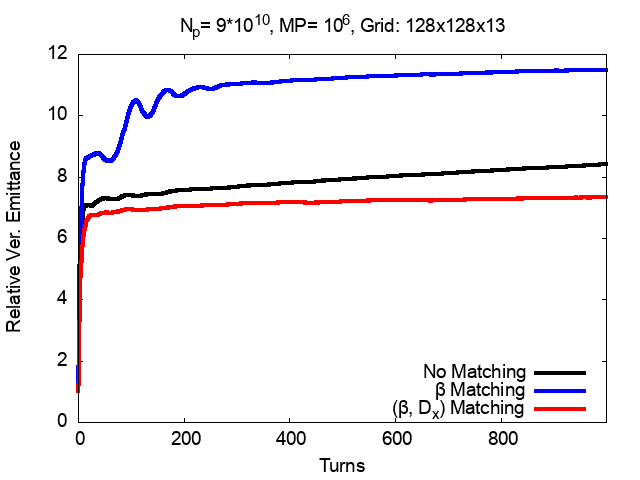}
\includegraphics[scale = 0.40]{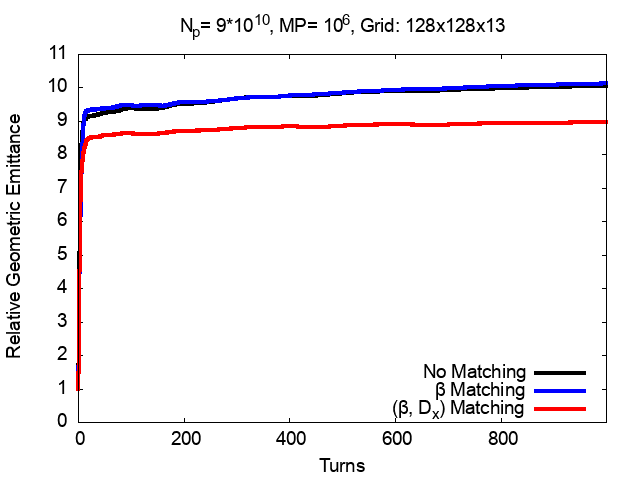}
\includegraphics[scale = 0.40]{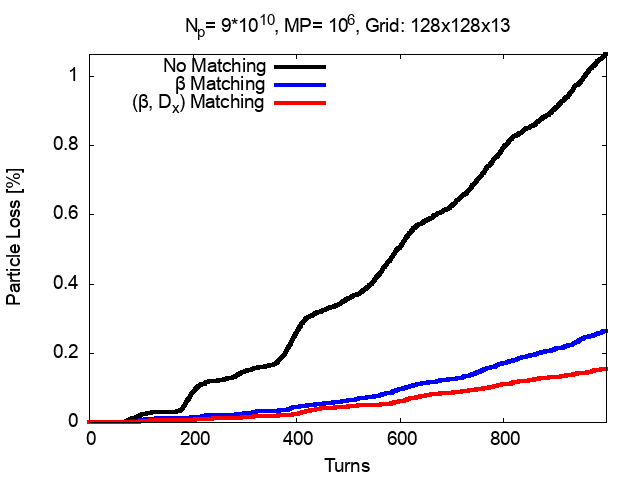}
\caption{Top row: Relative emittance growth in the horizontal (left) and vertical planes (right) with and without
  rms matching. Bottom row: The left plot shows the growth of the geometric  mean emittance
  $\sqrt{\eps_x \eps_y}$, the right plot shows the particle loss. }
\label{fig: emit_loss_match}
\efig
Fig. \ref{fig: emit_loss_match} shows the relative emittance and particle loss without matching and rms
matching just the $(\bt, \al)$ functions and also including the matching the horizontal dispersion $D_x$.
We observe that the $\eps_x$ decreases and $\eps_y$ increases with only beta matching relative to the values
without matching. Both these emittances decrease when we add $D_x$ matching. The left plot in the bottom
row shows that there is no change in the geometric mean emittance $\sqrt{\eps_x \eps_y}$ with only
beta matching while it drops slightly with additional $D_x$ matching. The right plot in the bottom row shows,
however, that just beta matching alone reduces the loss by about a factor of five and it drops another factor
of two with the addition of matching $D_x$. These results show that this rms matching does not significantly
affect the emittance growth nor does it slow the time scale of the initial increase.

\subsection{Slow Initialization}
\label{sec: slow_init}

 In this section we describe an alternative procedure to numerically reach a steady state.
Rather than injecting with the full charge,  the charge per  macroparticle is increased linearly from zero to its full value at
turn $T_{init}$, the initialization time. Provided this process is sufficiently adiabatic, one expects the beam to remain in
near equilibrium at every step as the beam has time to adjust to a slowly changing space charge force.

We note that in this procedure, the number of macro-particles stays constant while the charge per macro-particle
increases with time. This is easy to implement in a simulation since all the macro-particles are drawn from an initial
distribution; however, the procedure does not mimic the way bunch intensity increases in a real accelerator while the
machine is being filled. There are a
number of methods available that would allow to more realistically populate each bunch; the details of the  subsequent
dynamics will be influenced by the chosen method. Such an investigation is left for a future study.

The IOTA beam pipe radius is 25 mm; this value is used as the limiting aperture for simulation purposes. In the actual
machine, the available aperture in the nonlinear elements such as the octupoles or the nonlinear lens have smaller
apertures. These nonlinear elements are not included here.  To distinguish halo growth from core emittance growth, we
will also  discuss the dynamics with a larger aperture e.g. 100 mm. 
\bfig
\centering
\includegraphics[scale = 0.42]{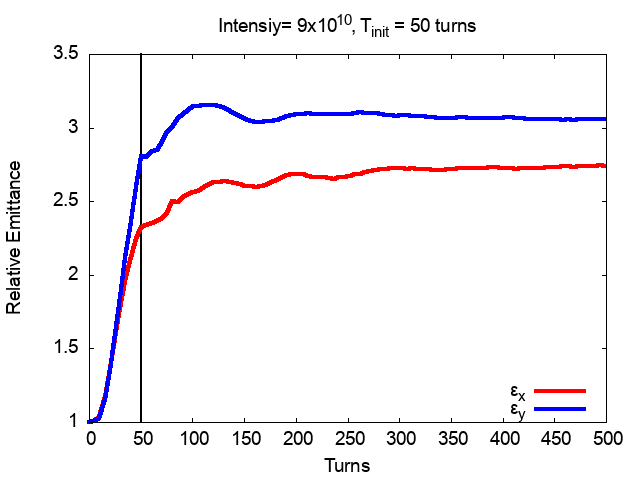}
\includegraphics[scale = 0.42]{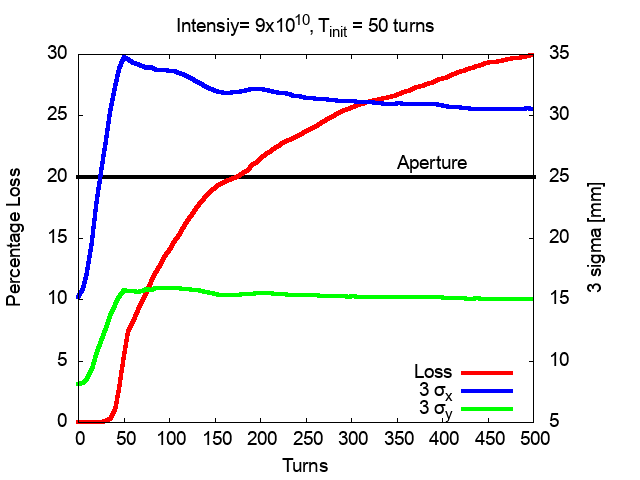}
\caption{Emittance growth and losses at a slow initialization time of 50 turns and with an aperture of 25 mm.}
\label{fig: Tinit_50}
\efig
Figure \ref{fig: Tinit_50}  shows emittance growth and beam loss for two choices of $T_{init}$
with the physical aperture set to 25 mm. Losses start when 3 $\sigma_x$ reaches the limiting aperture radius.
\bfig
\centering
\includegraphics[scale = 0.6]{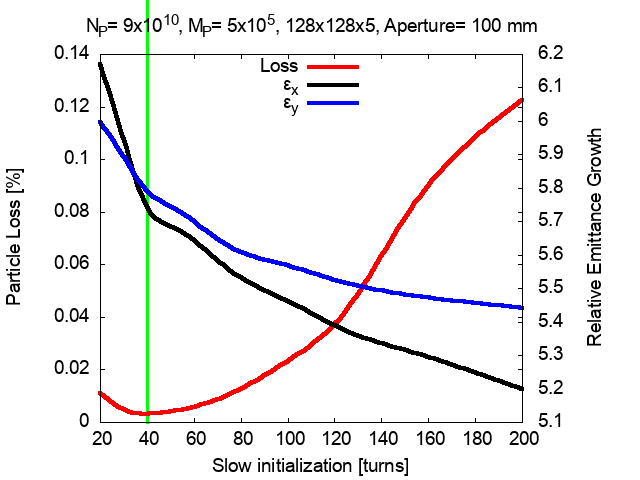}
\caption{Loss and emittance growth as a function of the slow initialization time.}
\label{fig: loss_emt_ninit}
\efig
Figure \ref{fig: loss_emt_ninit} shows emittance growth and particle loss after 500 turns as a function of the
slow initialization time. We observe that the  emittance growth decreases monotonically as $T_{init}$ increases and
falls    by about 15\% in the range studied. The losses
  increase by about 40 fold from the minimum to maximum over this range and are significantly more sensitive to
  the initialization time. We also observe that there is no reduction in the rms emittances over this range of $T_{init}$,
  so clearly   the core does not reach the aperture.
  These observations suggest that the halo is strongly affected by the charge on the macroparticles while the  growth
  of   the core is less affected. It may be useful to determine why the 
  loss is minimum at 40 turns. We note in passing that the relative emittance change in the transverse planes are nearly equal
  at this value of $T_{init}$, so that equipartitioning of emittances is apparently correlated with a minimum of losses. 
  \bfig
  \centering
  \includegraphics[scale = 0.42]{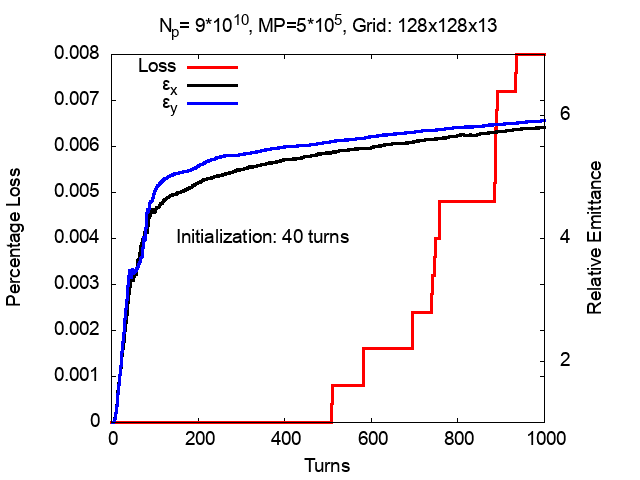}
  \includegraphics[scale = 0.42]{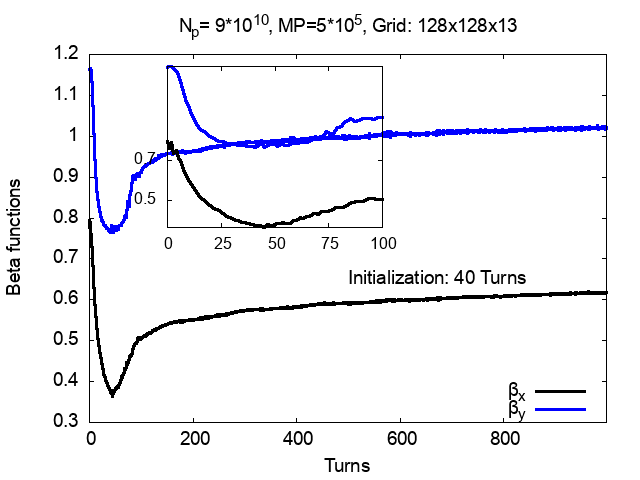}
  \includegraphics[scale = 0.42]{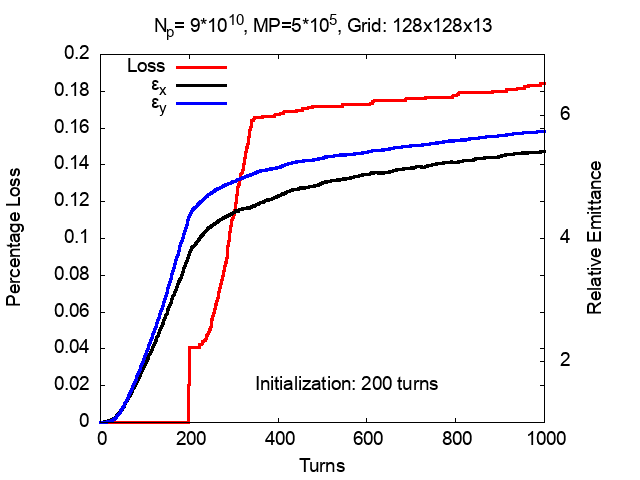}
  \includegraphics[scale = 0.42]{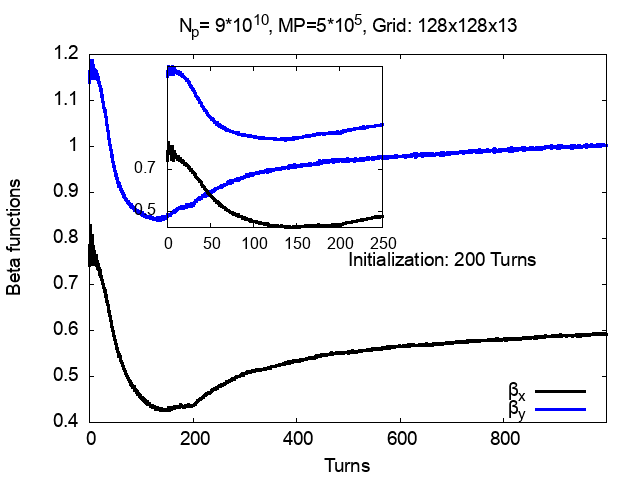}
  \caption{Left: Particle losses (left vertical scale) and relative emittance (right vertical scale), Right: Statistical
    beta functions for two initialization times : 40 turns (Top) and 200 turns (Bottom).}
  \label{fig: ninit_40_200}
  \efig
  \bfig
  \centering
\includegraphics[scale = 0.42]{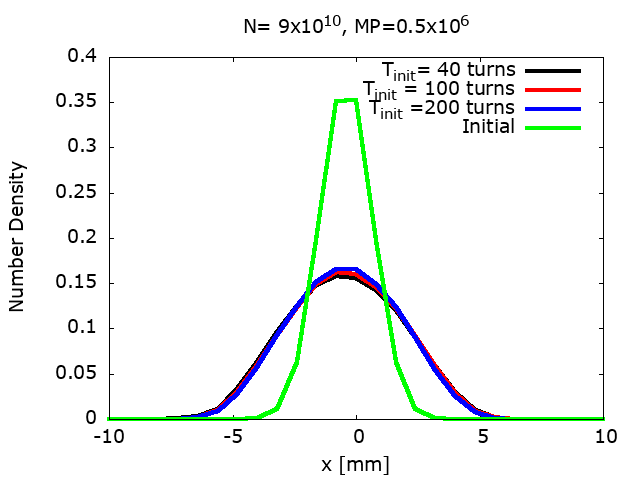}
\includegraphics[scale = 0.42]{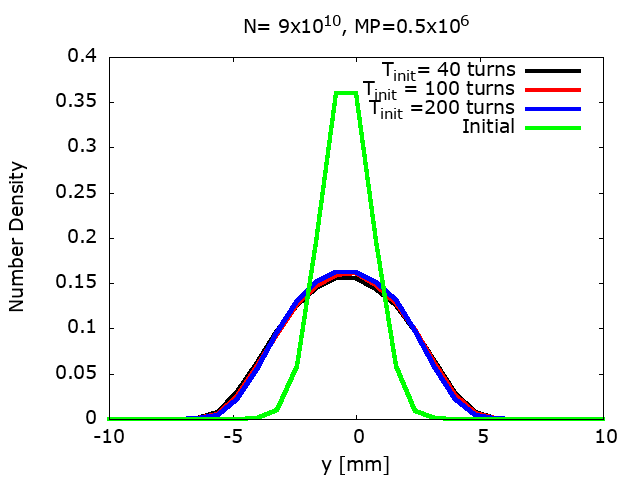}
\includegraphics[scale = 0.5]{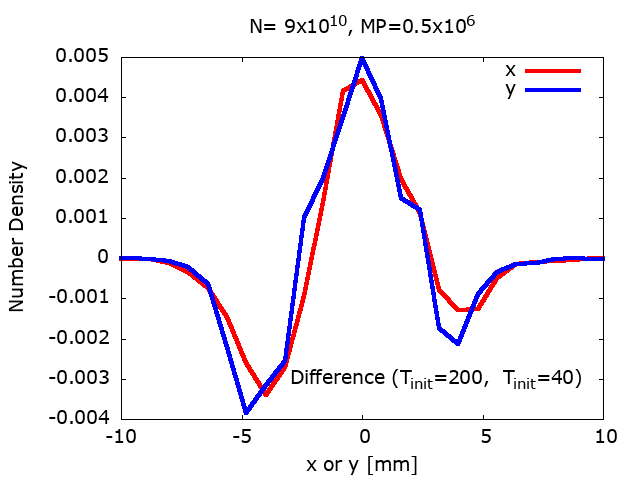}
\caption{Top: Density profiles for the  initialization times that correspond to the smallest and largest losses;
horizontal (left) and vertical (right). Bottom: difference in the profiles. }
\label{fig: profiles_ninit}
\efig
The two left  plots in  Fig. \ref{fig: ninit_40_200} show the evolution of the emittance growth and particle loss for two
initialization times $T_{int} = 40, 200$ turns. In both cases, the vertical emittance grows slightly more than the
horizontal emittance but the change is about the same for both 40 and 200 turn initializations. The horizontal emittance
growth is slightly less at $T_{init} = 200$ turns. The losses are significantly smaller at 40 turns. The right plots in this
figure shows the statistical beta functions for the same two initializations. The inset plots show that both
$\lan \bt_x \ran,\lan \bt_y \ran $ reach different stable values depend on $T_{init}$ and that the time to stabilize
increases with $T_{init}$. The top plots in Fig. \ref{fig: profiles_ninit} shows the initial and final horizontal and
vertical profiles  while the bottom plot shows the difference profiles between the two initializations in both planes.
These plots show that after stabilization, the profiles do not change much with the initialization time.

%\clearpage

\subsection{Convergence Tests on simulation parameters}
\label{section:num_test}

There are three  parameters in the pyORBIT simulations that need to be tested for convergence : the number of
macroparticles ($M_P$), the number of space charge kicks per betatron wavelength ($N_{sc}$) and the number of spatial
grid points ($N_x\times N_y\times N_z$) used while solving Poisson's equation.  Increasing $M_P$ reduces the statistical
noise and a larger $N_{sc}$ ensures better sampling of the spatially dependent space charge kick as the transverse beam
sizes vary along the ring. 
 These two parameters have practical upper limits set by the computing time required.  Increasing the number of grid points improves the resolution of the sampling of the space charge force.  An insufficient number of grid points
 increase the numerical noise with a PIC code and can lead to artificial emittance growth. Similarly, due to
 round-off error accumulation, an excessive grid size can lead to the same problem.
 There is therefore an optimum sampling that provides low enough noise \cite{Kesting_2015} for reasonable computational
 cost.  The number of grid points in each plane is distributed from
 $ - 3\sg_u$ to $3 \sg_u$, $u= x, y, z$ for Gaussian distributions in these planes,
although about 1\% of the macro-particles may be at larger amplitudes.

We describe tests performed to determine parameters that ensure 
convergence by examining the behavior of beam losses and emittance growth. The convergence tests are performed at full intensity  ($9 \times 10^{10}$) and we use the slow initialization procedure described above with $T_{init}=40$ turns.

pyORBIT assigns at least one space charge kick to each element so there are at least a hundred space charge kicks
around the IOTA ring. The exact number of space charge kicks per element is determined by the ratio $L/\Dl s$ where
$L$ is the element length and $\Dl s$ is a step length input parameter. Summing over the elements and dividing by the
integer part of the tune, $\sim 5$ in IOTA yields the number of space charge kicks per betatron period, $N_{SC}$.
 For example, $\Dl s = 0.1$m yields $N_{SC} = 119$ while at a lower resolution
choosing $\Dl s = 0.4$m yields $N_{SC}$ = 56.

\bfig
\centering
 \includegraphics[scale = 0.42]{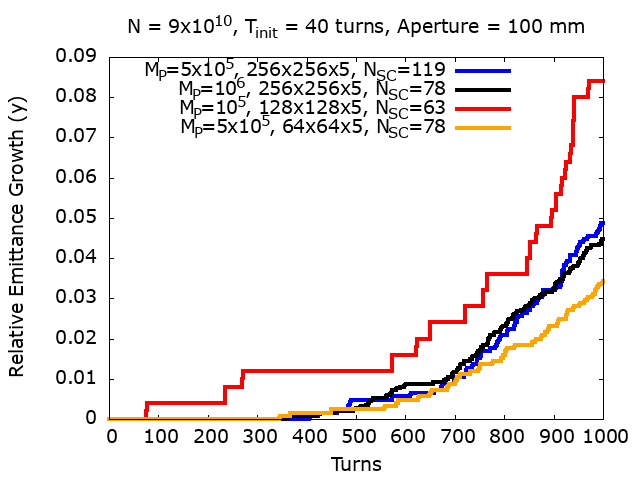}
 \includegraphics[scale = 0.42]{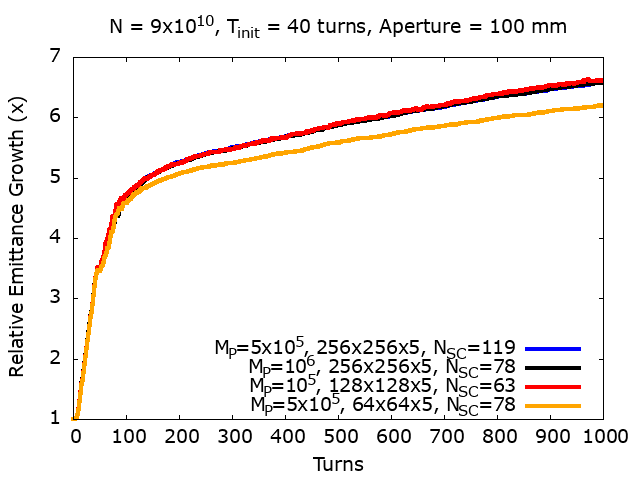}
 \caption{Particle losses (left) and horizontal emittance growth (right) for different combinations of macro-particle
   numbers,   grid sizes and number of space charge kicks. The vertical emittance has the same dependence  as the
 horizontal emittance on these parameters. }
 \label{fig: all_losses_emit}
 \efig
 Figure \ref{fig: all_losses_emit} shows the particle loss and emittance growth for different  numbers of
 macro-particles, grid sizes and space charge kicks.  Other parameters that are held constant are shown at the top of
 each  plot.  We observe that losses are generally more sensitive to changes in parameters than the emittance. The losses plot shows that  $10^5$ macro-particles and a grid size of $64\times 64 \times 5$ insufficient. The emittance plot shows that $10^5$ macro-particles suffices while a $64 \times 64 \times 5$ grid
 yields a lower emittance than finer grids. 

\bfig
 \centering
 \includegraphics[scale = 0.60]{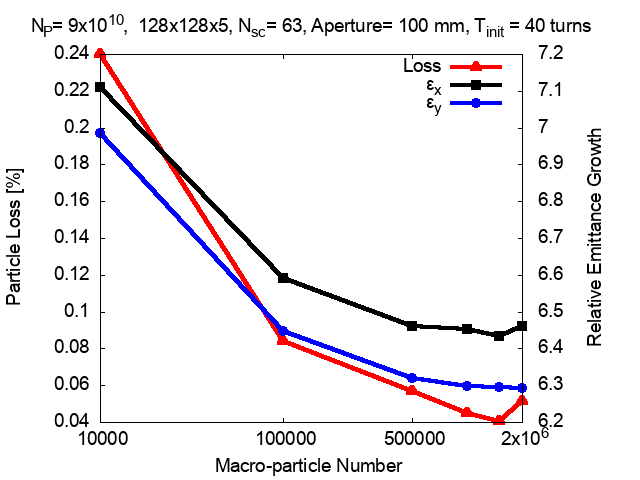}
 \caption{Particle loss (red; left vertical  scale) , relative emittance growth  (black, blue; right vertical scale) after 1000 turns as a function of  the
   macro-particle number varying over three orders of magnitude,shown on a log scale.  The constant parameters are
 shown at the top.}
 \label{fig: loss_exy_MP}
 \efig
 Fig. \ref{fig: loss_exy_MP} shows the dependence of particle loss and emittance growth  on the macro-particle
 number $M_P$ with the number of grid points fixed at $128 \times 128 \times 5$ and
 number of space charge kicks $N_{sc}$ per betatron wavelength fixed at $63$.
 Over the range $10^3 \le M_P < 5\times 10^5$, the loss is reduced by about a factor of  four while
 at larger values of $M_P \ge 5 \times 10^5$ the  loss changes by less than 0.05\%. The emittance stays nearly
 constant over this latter range from which we conclude that $M_P = 5 \times 10^5$ is the minimum number of
 macro-particles required. 
 \bfig
  \centering
 \includegraphics[scale = 0.60]{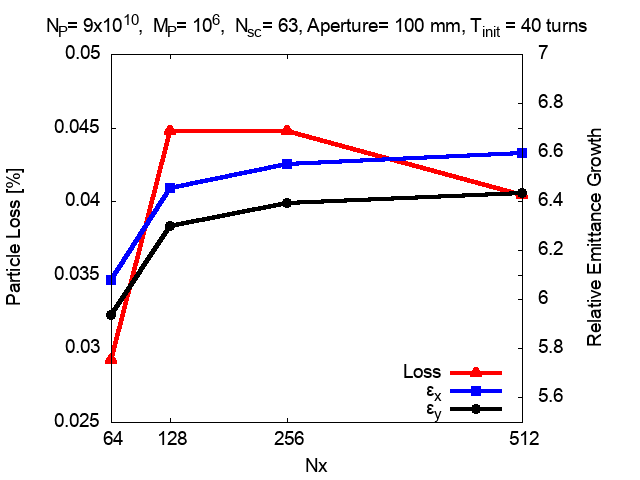}
 \caption{Particle loss (red) , relative emittance growth  (black, blue) after 1000 turns as a function of  the
grid parameter $Nx = Ny$, $Nz = 5$.  The constant parameters are shown at the top.}
 \label{fig: loss_exy_Nx}
 \efig
Fig. \ref{fig: loss_exy_Nx} shows the variation of the loss and emittance growth  with the number of grid points
$N_x = N_y$ with the number of macro-particles fixed at 10$^6$ and  $N_{sc}$
fixed at $78$.  The emittance stays nearly constant for $ 128 \le N_x \le  512$ and 
the loss fluctuates by less than 0.005\% over this range. 
This   shows that 128 is the minimum number of grid points required.
 \bfig
  \centering
 \includegraphics[scale = 0.60]{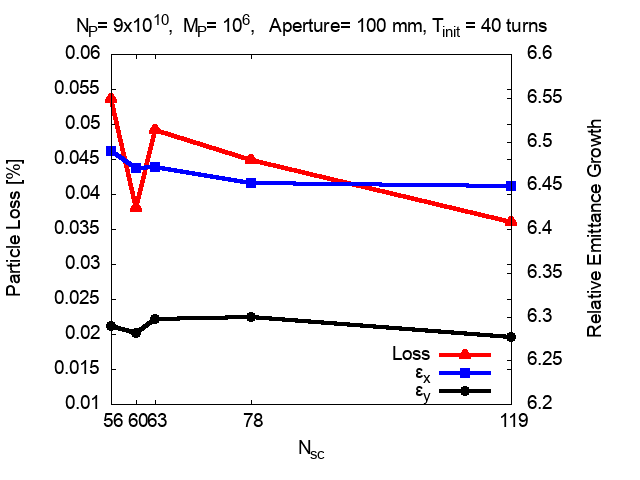}
 \caption{Particle loss (red) , relative emittance growth  (black, blue) after 1000 turns as a function of  the
   number of space charge kicks $N_{SC}$ per betatron period  The constant parameters are shown at the top.
The values of $N_{SC}$ correspond to the step length parameter $\Dl s = 0.1, 0.2, 0.3, 0.4$m discussed in the text. }
 \label{fig: loss_exy_int}
 \efig
 Fig. \ref{fig: loss_exy_int} shows the convergence with respect to $N_{SC}$ with the macro-particle number
 $M_P=10^6$ and grid size 128x128x5 held constant. We observe that the loss fluctuates by $\sim 0.015$\% while
 the emittances are nearly constant with $N_{SC} \ge 63$. 

\section{Small amplitude tune shifts and tune footprints }

The goal of this exercise is to validate the accuracy of pyORBIT's space charge model by comparing the 
 tune shift obtained by tracking to analytical predictions. As mentioned earlier, 
the only nonlinearity in the lattice model arises from space charge.

It has been pointed out \cite{Schmidt_2014} that chaotic motion is observed for small amplitude particles in PIC
codes due to numerical noise. One can expect that the effects of this chaos to mostly disappear when ensemble
averaging over many
particles to calculate statistical variables such as emittances, beam sizes etc. There are likely few analytical estimates of
emittance growth with nonlinear space charge.

Keeping in mind the above caveat for tunes at small amplitudes, tune shifts as a function of amplitude due to space charge 
can nevertheless be compared to analytical results for a Gaussian distribution.

The analytical small amplitude tune shift for a transverse    Gaussian distribution is:
\begin{equation}
    \Delta Q_{0, sc} = \frac{r_p}{\beta_k \gamma^2 \epsilon_N} \bar{\lambda}_L N_p R
\end{equation}
where $r_p$ is the classical proton radius, $\beta$ and $\gamma$ are Lorentz factors, $R$ the effective machine
radius, $\epsilon_N$ the normalized transverse emittance, $N_p$ the number of protons per bunch, and
$\bar{\lambda}_L$, the effective line density. In all tests, the reference  particle at $(z=0, dE=0)$ does not move.

The exact position dependent line density $\lm(z)$ for a longitudinal water-bag distribution and the
average effective  line density $\bar{\lm}_L  $ are given by \cite{Sen_unpub}
\beqr
\lm(z) & = & \fr{1}{C_{wb}} \sqrt{ \cos( k_{rf}z ) - \cos( k_{rf} z_{max}) }  \label{eq: lambda_N} \\
\bar{\lm}_L & = & \fr{1}{2 C_{wb}} \sqrt{1 - \cos( k_{rf} z_{max})}   \\
  C_{wb} & =  & \fr{4}{k_{rf}} \sqrt{1-\cos (k_{rf} z_{max})} E\left(\half k_{rf} z_{max} |  \csc ^2(\half k_{rf} z_{max} \right)  
\eeqr
Here $k_{rf}$ is the rf wave-number $z_{max}$ is half the total bunch length and $E$ is the
complete elliptic integral of the second kind.  In the limit of short bunches where $k_{rf}z_{max} \ll 1$
the water-bag distribution reduces to the elliptic distribution with a normalized density
\beq
\lm(z) = \fr{2}{\pi z_{max}} \sqrt{1 - (\fr{z}{z_{max}})^2}
\eeq
The IOTA bunch is long and fills the bucket, so the elliptic distribution is not  a good approximation. 

For the water-bag distribution with $z_{max}= 4.5$ meters, the effective line density  $\bar{\lm}_L = 0.083$.
Plugging in all known parameters, and accounting for the fact that $\epsilon_N$ changes during tracking we
calculate the small amplitude tune shift:
\begin{equation}
    \Delta Q_{0, sc} = \frac{-1.09*10^{-17} * N_p} {\langle \epsilon_N \rangle}
    \label{dQ_eq}
\end{equation}
where $\langle \epsilon_N \rangle$ denotes a suitably time-averaged emittance. We find that the transverse tune
  shift does not depend very sensitively on the longitudinal distribution. Figure \ref{fig: dQ_zmax} shows the absolute value of the space charge tune shift as a function of the maximum longitudinal extent $z_{max}$ of a bunch for different distributions. At $N_b = 9\times 10^{0}$ and $z_{max}= 4.5$m, the tune shifts are
$\Dl Q_{sc, 0} = (-0.554, -0.521, -0.514)$ for the Gaussian, parabolic and water-bag distributions respectively. 
\bfig
  \centering
  \includegraphics[scale = 0.5]{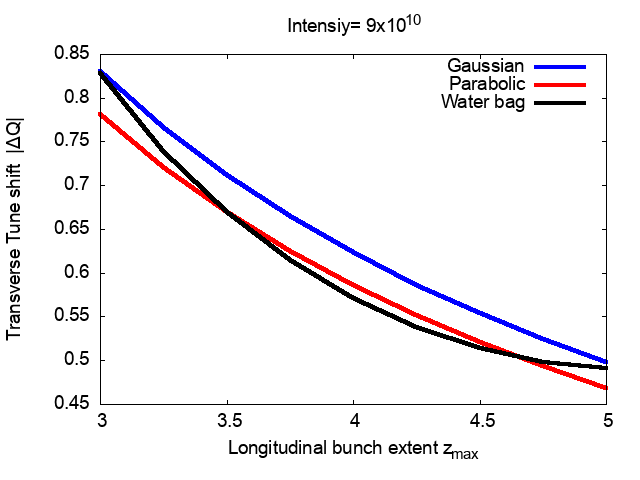}
\caption{Absolute transverse tune shift as a function of the longitudinal extent $z_{max}$ for different distributions. }
  \label{fig: dQ_zmax}
  \efig

\subsection{Small Amplitude Tune Shift}
\label{0_amplitude_test}

We calculate the transverse tunes using two different methods.   First we use the definition of the tune as the number
of transverse oscillations in one revolution to calculate the small  amplitude tune. 
The bunch is initialized as shown in Figure \ref{fig:distribution_example}. It is tracked for 500 turns and the first
200 turns are used for slow initialization. At this point, a test particle is added at a nearly zero amplitude
($0.01\sigma_x$, 0, $0.01\sigma_y$, 0, 0, 0). The bunch is tracked for an additional 1000 turns, during which all
monitors  in the IOTA lattice record the transverse position of the test particle at every turn. Counting the total
number of zero crossings and dividing by twice the number of turns monitored yields a lower bound for the
small amplitude tune.
\begin{figure}%%%
  \centering
  \includegraphics[scale = 0.15]{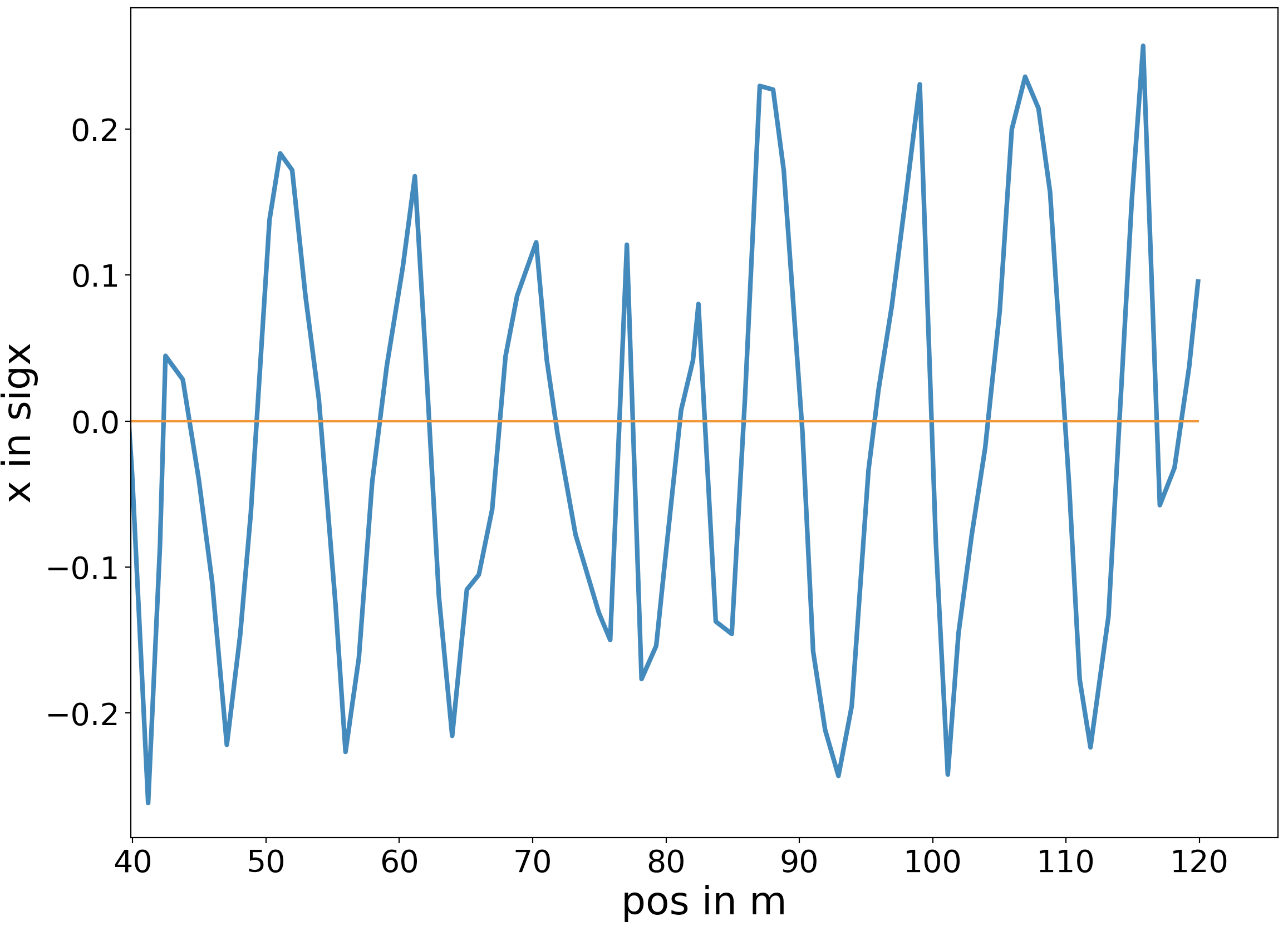}
    \includegraphics[scale = 0.15]{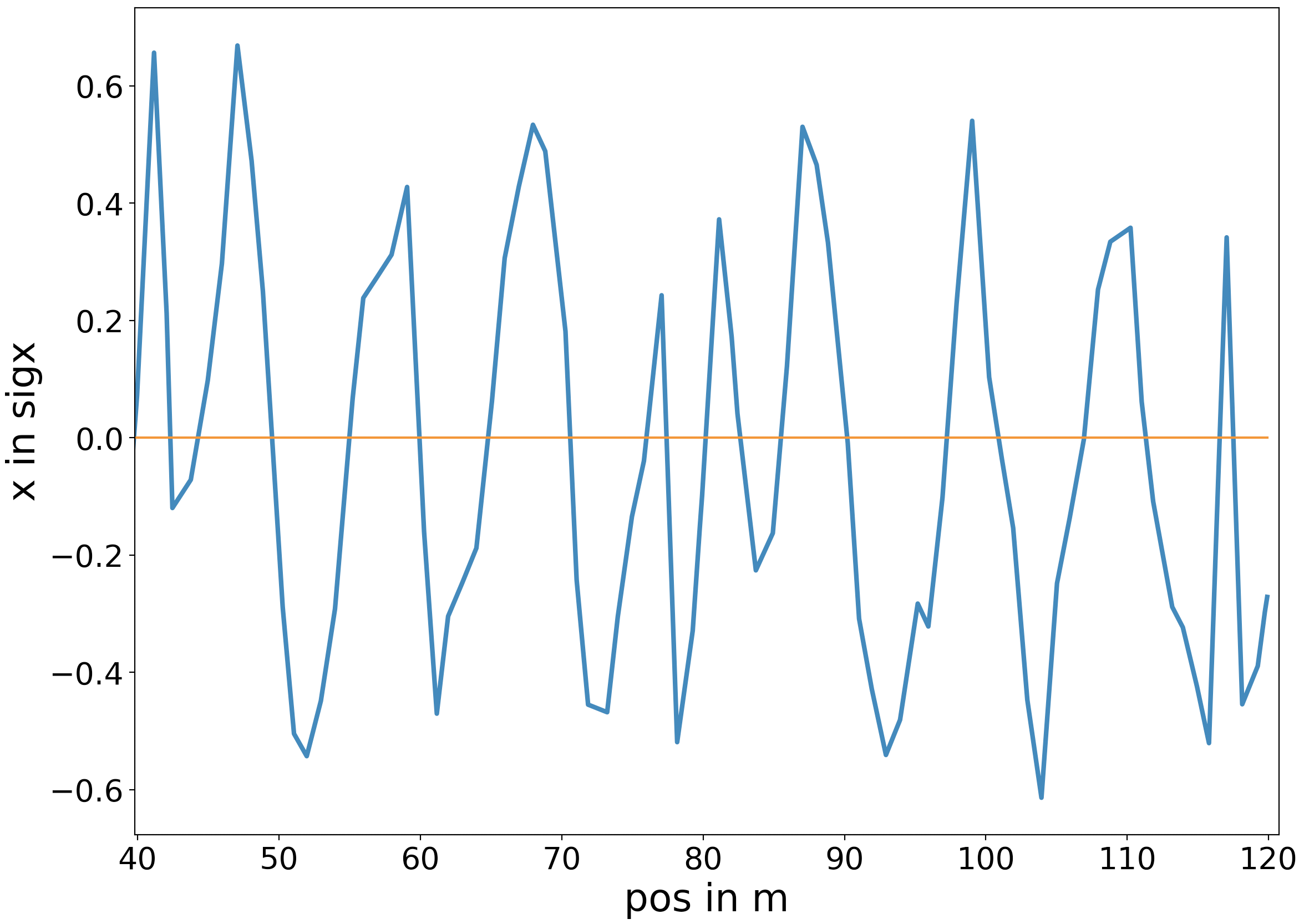}
    \caption{Betatron oscillations over two turns : (a) Zero intensity (left), 21 zero crossings
 (b) Bunch charge $9 \times 10^{10}$ (right), 19 zero crossings. }
    \label{fig: beta_osc}
    \efig
    Fig. \ref{fig: beta_osc} shows the betatron oscillations over two turns for two bunch intensities.
     The number of zero crossings  allows us to extract the tune with a precision of 0.25 over  the following ranges:
     (a) Zero and      $10^9$ intensities: $5.25 < Q_x < 5.5$, (b) $10^{10}$ intensity: $5.0 < Q_x < 5.25$
     and (c)  $10^{10}$ intensity: $ 4.75 < Q_x < 5.0$. Table \ref{table: 0amp_tunes} shows the tunes
    calculated from betatron oscillations with higher precision using 1000 turns and the fractional tunes calculated
    with an FFT using the same number of turns. The FFT tune is averaged over 100 particles placed at the same small
    amplitude in an effort to reduce the fluctuations due to chaotic motion reported in \cite{Schmidt_2014}. 
\begin{table}
  \bec
  \btable{|c|c|c|} \hline
  Intensity & \multicolumn{2}{|c|}{ Small amplitude tunes $Q_x, Q_y$} \\
  & Betatron Oscillations & FFT \\ \hline
  0  & (5.300, 5.301)  &  (0.2988, 0.3) \\
  $10^9$ &  (5.2681, 5.267) & (0.2663, 0.265) \\
  $10^{10}$ & (5.037, 5.023)   &  (0.039, 0.035)  \\
  $9 \times 10^{10}$ & (4.873, 4.973)   &  (0.951, 0.975)   \\
  \hline
  Intensity  &  \multicolumn{2}{|c|}{Small amplitude tune shifts} \\
    &  Theory  &  Simulation  \\ \hline
 0  &  (0.0, 0.0) &  (0.0, 0.001)   \\
10$^9$  & (-0.037,  -0.037) &  (-0.03, -0.034)   \\
10$^{10}$  &   (-0.29, -0.29) &  (-0.262, -0.266)  \\
9$\times 10^{10}$         &  (-0.514 -0.514) & (-0.50, -0.536) \\
\hline
\etable
  \caption{Small amplitude tunes from betatron oscillations and the fractional tunes from an FFT, in both cases
    using 1000 turns of data.  Also shown are the small 
 amplitude transverse tune shifts  calculated from theory and simulations   at different intensities.}
  \label{table: 0amp_tunes}
\eec
\end{table}
\subsection{Tune Footprints }

The amplitude-dependent tunes for a transverse Gaussian distribution can be calculated analogously to those from a
head-on beam-beam interaction between two Gaussian beams. For a round beam, the tunes are
\cite{Sen_unpub}
\beqr
\Dl Q_{x, sc} & = & \Dl Q_{0, sc} \int_0^{\infty} \exp[-(\al_x+\al_y)u] I_0 (\al_y u)  \left[ I_0( \al_x u) - I_1( \al_x u) \right] du  \label{eq: dQx_SC} \\
\Dl Q_{y, sc} & = &  \Dl Q_{0, sc} \int_0^{\infty} \exp[-(\al_x+\al_y)u] I_0 (\al_x u)  \left[ I_0( \al_y u) - I_1( \al_y u) \right] du  \label{eq: dQy_SC} \\
\al_x & = & \fr{a_x^2}{4}, \;\; \;\; \al_y = \fr{a_y^2}{4}, \;\; \;  a_x = \fr{x}{\sg_x}, \; a_x = \fr{y}{\sg_y}
\eeqr
where $ a_x, a_y$ are the dimensionless amplitudes. Slightly more complex expressions exist for
non-round beams, but those will not be used here.

\begin{figure}%%
  \centering
  \includegraphics[scale = 0.425]{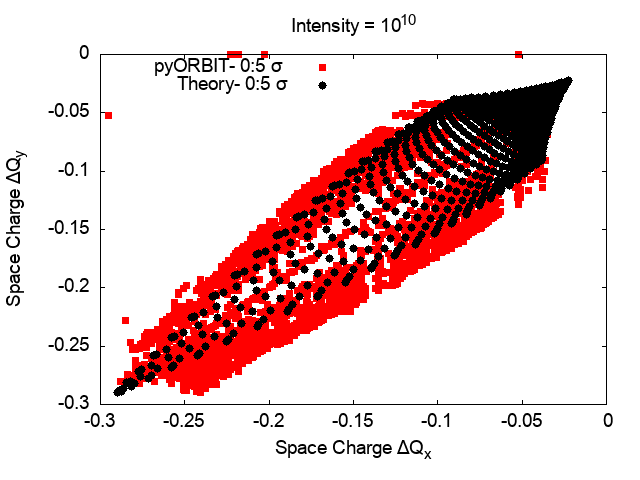}
  \includegraphics[scale = 0.425]{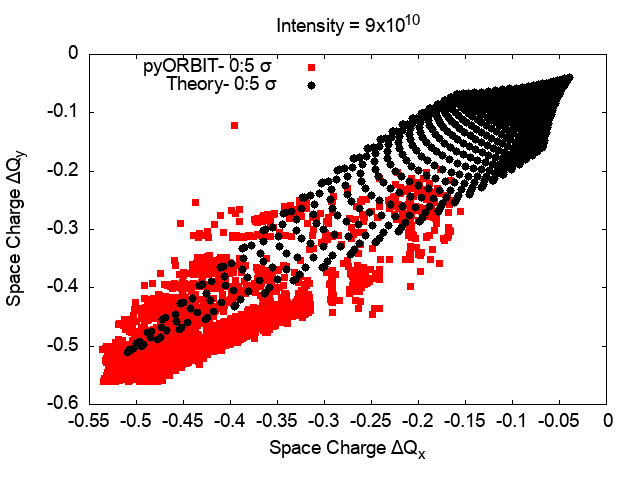}
  \caption{ Tune shifts with amplitude due to space charge  from pyORBIT and theory at intensity $10^{10}$ (left)
intensity $9\times 10^{10}$ (right). In both cases, the pyORBIT simulation ranges from 0-5$\sg$ (due to aperture
limitations) and the theoretical footprint using Eqs.(\ref{eq: dQx_SC}) and (\ref{eq: dQy_SC}) extends to 5$\sg$.
}
    \label{fig:tune_spec_10}
\end{figure}
The tune footprints are calculated with a bunch initially tracked for 1000 turns (including 40 turns of slow
initialization, as described in Section \ref{sec: slow_init}) to allow emittance to stabilize
and get closer to an equilibrium value. After this initial period, 5000 test particles are injected with initial conditions
distributed along semi-circular arcs on radii from (0-5)$\sg$ in $x-y$ space; particles initially at larger amplitudes would
be lost at the physical aperture. The bunch and the test particles are tracked for another 1000 turns to obtain the
fractional tunes using an FFT. 
 At intensity $N_b= 9\times10^{10}$,  the lattice tune is shifted to $Q_x = 5.46$ and
 $Q_y = 5.44$ to avoid the presence of resonances (possibly ``stochastic resonances'') within the footprint. The tune
 footprints  in Figure \ref{fig:tune_spec_10} show both theoretical and pyORBIT simulation results. The theoretical
 footprint also extends to 5$\sg$. At intensities at below 10$^{10}$
 and below, the simulated footprint matches the theoretical footprint quite well but is wider.  At 9$\times 10^{10}$,
 the asymmetry between $\Dl Q_{x, sc}, \Dl Q_{y, sc}$ in the simulated tunes increases with
 $\mid \Dl Q_{x, sc}\mid < \mid\Dl Q_{y, sc}\mid$, possibly due to the dispersion contribution to the horizontal beam
 size. 
 Simulations of the emittance growth (discussed in Section \ref{sec: emit_loss}) show that the final emittance is
 about 6.4 times larger than the initial emittance; this was the value assumed for $\eps_N$ in the theoretical footprint.
 The small amplitude tuneshift in Table \ref{table: 0amp_tunes} shows that the difference between the average of the
 simulated tune shifts $(\Dl Q_{x, sc} + \Dl Q_{y, sc})/2 $ and the theoretical value is 0.004, a few times 0.001, the
 precision of the FFT calculation; however, the simulated footprint is significantly wider (by about 0.05 in tune units)
 than the theoretical footprint. At large amplitudes close to 5$\sg$, the simulated tune shifts are larger than expected
 from theory.  Some differences are to be expected since the theoretical model assumes a zero length bunch, so does
 not include the longitudinal dependence of the transverse space charge force. Nor does the theory include the effect of
 dispersion but this can be done easily, if required. Finally there is numerical noise in the PIC code which will impact the
tune  footprint which is  calculated without averaging. 
%---------------------
\section{Conclusions}
%---------------------
\bit
\item 
  The symplectic nature of single particle tracking with pyORBIT was checked for both the linear lattice and the
  lattice with octupoles using two measures: $(\det \mtx{J} - 1)$ and $ \mathrm{max}(\mtx{J}^T  \mtx{S} \mtx{J} - \mtx{S})$.
  $\mtx{J}$ is the Jacobian matrix of the transfer map and $\mtx{S}$ is the symplectic matrix. On both
  measures, the pyORBIT model's  deviation from symplecticity is $ \sim 10^{-11}$ for the linear lattice.
  This deviation increases to $\sim 10^{-4}$ with the octupoles. By comparison, the same tests using MADX
  show the same measures to be $\sim 10^{-8}$ for the linear lattice and $\sim 10^{-4}$ with octupoles. 

  The variance of the Hamiltonian is similar with both codes.

  The transverse tunes calculated with octupoles in the lattice agree to within 10$^{-3}$ over the stable region
  of phase space. The boundary where the differences are about an order of magnitude larger is close to the
  dynamic aperture; the tune difference map resembles an FMA map produced by a single code.  The
  dynamic apertures  found by the two codes  agree remarkably to within 1\% with or without synchrotron
  oscillations (see Table \ref{table: DA_oct} ).

\item 
  The main purpose of using pyORBIT is for its space charge tracking which is based on a PIC style method.
  Such codes are not symplectic and indeed find that the deviations from symplecticity are about two orders
  of magnitude larger than in the single particle version with octupoles. This shows that the space charge
  tracking results cannot be used with confidence for long-term tracking. However we expect that short term
  results ($\sim 10^3 - 10^4$ turns) should be reliable for parametric studies.

\item 
  The IOTA lattice we have used for the space charge studies is the original linear lattice. This lattice is not rms
  matched
  to the space charge lattice functions and the mismatch increases with the bunch intensity. At full intensity,
  this mismatch leads to almost ten-fold increase in transverse emittance and the beam loss is about 1\%
  after 1000 turns. Most of the emittance increase occurs over the first few turns due to a large mismatch.
  In order to reduce the emittance growth and loss in this mismatched lattice, first we matched the lattice
  for a Gaussian distribution with initial rms sizes the same as the final values. This reduced the emittance growth
  slightly   but lowered the losses to 0.1\% over 1000 turns. Next   we introduced a slow
  initialization time $T_{init}$ (without rms matching)  during which the charge on each macro-particle is increased
  linearly over time.
  We find that this reduces both the emittance growth to about five-fold and especially the losses to less than
  0.1\%  over the same time.   The optimum initialization time  $T_{init}$ is found to be 40 turns for which
  the loss is about 0.01\%.

\item 
  With the optimum slow initialization time fixed at 40 turns, we tested for convergence by examining the
  behavior of the emittance growth and the particle loss as one of three important parameters was varied
  while keeping the other two parameters fixed. In general, the loss shows more fluctuations than the
  emittance as the parameters are varied. However with the aperture set at 100 mm (to avoid an emittance
  reduction due to significant losses), the losses are typically around 0.05\% and fluctuations at the 10\%
  level are not significant. 
  Figure \ref{fig: loss_exy_MP} shows the minimum number of macro-particles needed is 5$\times 10^5$
  keeping the grid size fixed at 128x128x5 and the number of space charge kicks per betatron wavelength
  $N_{sc}=63$. Figure \ref{fig: loss_exy_Nx} shows that the emittance growth stabilizes at the
  grid size 128x128x5 and does not change much upto grid sizes 512x512x5. Finally
  \ref{fig: loss_exy_int} shows that convergence is achieved with $N_{sc}=78$ but $N_{sc}= 63$ can
  be used without significant differences. This value of $N_{sc}$ is quite a bit larger than typically expected. 

\item
   The space charge small amplitude tune shifts and tune footprints agree quite well with theoretical predictions
  up to a bunch intensity of $10^{10}$. At the maximum intensity of $9 \times 10^{10}$, the small amplitude
  tune shifts still agree reasonably but the simulated footprint is somewhat wider than the theoretical
  footprint (see Fig \ref{fig:tune_spec_10}b). In making the comparison, we have to take into account the
  emittance growth at full intensity. 
  \eit

  To sum up, we have established that pyORBIT matches MADX very closely for single particle tracking. The space charge
  model of pyORBIT is not designed to be symplectic, so naturally the deviations from symplecticity are significant.
  Nevertheless, we find that the space charge tuneshifts calculated over 1000 turns  with pyORBIT agree
  reasonably well with theory. We believe therefore that over the time scale of 10$^3$- 10$^4$ turns, pyORBIT can be
  used with confidence. This report dealt only with the transverse space charge effects, so similar issues will be need
  to be examined when the longitudinal space charge effects are included.

  \noi   {\large \bf Acknowledgments} \\
  We thank the Lee Teng internship program at Fermilab which enabled the participation of Runze Li in this study.
Fermilab is operated by Fermi Research Alliance LLC under DOE contract No. DE-AC02CH11359.

\clearpage 

\appendix

%\section*{Appendices}
\addcontentsline{toc}{section}{Appendices}
\renewcommand{\thesubsection}{\Alph{subsection}}

\subsection{Appendix: Beam temperature and chromaticity}
\label{section: appendix_A}
\setcounter{equation}{0}
\renewcommand{\theequation}{A.\arabic{equation}}

The transverse temperature is
\beq
T_{\perp} = \gm m_0 \lan v_{\perp}^2 \ran
\eeq
The mean squared transverse velocities  can be found from
\beqrs
v_{\perp} & = &  \fr{p_{\perp}}{m_0} =  \fr{p_{0}}{m_0}u', \;\;\;\; u = x, y  \\
  \lan v_x^2 \ran & = & (\fr{p_0}{m_0})^2 \lan (x')^2 \ran = (\fr{p_0}{m_0})^2 \gm_x \eps_x \\
  & = & \frac{1}{(m_0 c)^2}[E^2 - (m_0 c^2)^2] \gm_x \eps_x  =  c^2[\gm^2 - 1] \gm_x \eps_x 
  \eeqrs
  The above average represents an average over the beam distribution.

Averaging over the ring, we have
  \beq
\lan k_B  T_{\perp} \ran =  m_0 c^2 [\gm^2 - 1] \eps_{x, N} < \gm_x >
  \eeq
    To find the ring average of $\gm_x$, we use \cite{Syphers}
  \beqr
  K_x \bt_x & = & \al_x' + \gm_x  \\
  \Rarw -4 \pi Q_x' & = & \oint (\al_x' + \gm_x) ds = \oint \gm_x ds = C \lan \gm_x \ran
  \eeqr
  where $K_x$ is the quadrupole gradient parameter,  $Q_x'$ is the natural chromaticity and $C$ is the machine circumference. Hence
  \beq
\lan  \sg_{x'}^2  \ran  = \lan \gm_x \ran \eps_x  = - \fr{2 Q_x'}{R} \eps_x  \label{eq: div_rms}
  \eeq
where $R$ is the machine radius. We recall that the average rms size is given approximately by
  \beq
\lan  \sg_x^2 \ran = \lan \bt_x \ran \eps_x = \fr{R}{Q_x} \eps_x   \label{eq: sig_rms}
\eeq
We note that while Eq. (\ref{eq: sig_rms}) for the rms size requires the smooth focusing approximation wherein
$\lan 1/\bt_x \ran  \approx 1/\lan \bt_x \ran $, Eq. (\ref{eq: div_rms})  for the divergence does not require
any approximation and remains valid even with nonlinear magnets in the ring. 
From these two equations we observe that  the average beam size and divergence are determined by the tune and
natural chromaticity respectively. 

The average temperature over the ring is therefore 
\beq
\lan k_B T_x \ran = m_0 c^2 (\gm^2 - 1) [\fr{-2 Q_x'}{R}] \eps_{x, N}
  \eeq
where $\eps_{x, N}$ is the  normalized emittance.

%\appendix
\subsection{Appendix:  Additions and  changes to pyORBIT}
\label{'section: appendix_b'}
\setcounter{equation}{0}
\renewcommand{\theequation}{B.\arabic{equation}}

This appendix describes additions and modifications to the original pyORBIT program\cite{pyORBIT}, including a new dipole edge element and the optional elimination of nonlinearities in the dipole and quadrupole elements. A number of bugs found in pyORBIT have been corrected. All additions and modifications are included in our repository on github \cite{pyORBIT}.

\subsubsection{Adding Dipole Edge Element in pyORBIT}

In the original version of pyORBIT, there is no distinct dipole edge element. In order to add dipole edge corrections to pyORBIT, changes were made in three sections of the code: the parser, the tracking method and the python element wrapper.

In the parser, found in the source file {\tt py-orbit/py/orbit/teapot/teapot.py} a new dipole edge element class named {\tt DipedgeTEAPOT} was introduced. When an element of type "dipedge" is encountered by the {\tt pyORBIT MADX} sequence parser, a new {\tt DipedgeTEAPOT} element is instantiated. Its parameters, {\tt e1}, {\tt h}, {\tt hgap}, and {\tt fint} correspond to the dipole edge parameters in {\tt MADX}. {\tt e1} is the rotation angle for the pole face, {\tt h} is the curvature $\frac{1}{\rho}$, {\tt hgap} is the half gap of the (upstream/downstream) bending magnet and {\tt fint} is the edge field integral. The edge is implemented as a thin element i.e. it has length 0. The {\tt DipedgeTEAPOT} tracking method invoked by the tracker is defined in the source file {\tt py-orbit/src/teapot/teapotbase.cc}. The implementation considers only first order terms. The corresponding transfer matrix is\cite{dip}:

\beqr
         \begin{bmatrix}
           x \\    x' 
         \end{bmatrix} & = & 
    \begin{bmatrix}
    1 & 0 \\       h  \tan(e) & 1\\
    \end{bmatrix}
              \begin{bmatrix}
           x_0 \\              x'_0
         \end{bmatrix}           \\ 
           \begin{bmatrix}
           y \\            y' 
         \end{bmatrix} & = & 
    \begin{bmatrix}
    1 & 0 \\      -h  \tan(e - \psi) & 1\\
    \end{bmatrix}
    \begin{bmatrix}
      y_0 \\               y'_0
         \end{bmatrix}       \\ 
\psi &  = &  {\tt fint} \cdot 2 \cdot {\tt hgap} \cdot h \cdot \frac{1 + \sin^2(e)}{\cos(e)}
\eeqr
where $e$ is either the entrance or the exit edge angle. 

A new wrapper method was added to make the edge element and its methods available from {\tt python}. The implementation can be found in the file \\
{\tt py-orbit/src/teapot/wrap\_teapotbase.cc}. 

{\tt pyORBIT} now handles the dipole edges in the IOTA lattice {\tt MADX} sequence. When these edges are accounted for, the bare lattice tune $Q_y$ calculated by {\tt pyORBIT} is 5.3, in agreement with {\tt MADX}; without them {\tt pyORBIT} predicts 5.42. 
\subsubsection{Nonlinear Effects of Magnet Edge}
{\tt MADX} and {\tt pyORBIT} use different approaches for tracking. {\tt MADX} operates converts every element to thin kicks. In contrast {\tt pyORBIT} treats some elements as thick lenses. These elements are further subdivided; a variable called {\tt nparts} controls the subdivision. For example, a dipole with nparts = n, is treated as follows by {\tt pyORBIT}:

\begin{center}
{\tt Fringe In, Entrance, Dipole\_1, ....., Dipole\_n-2, Exit, Fringe Out}
\end{center}

{\tt Fringe In} and {\tt Fringe Out} account for edge effects the other n parts (including entrance and exit) are main body of this dipole element. The length (and bending angle for dipole) is distributed among each of these n parts according to $l_{entrance} = l_{exit} = \frac{l_{others}}{2} = \frac{L}{2*(n - 1)}$. The entrance contains only linear transformation and other elements have both linear and nonlinear part. 
Since every element needs an entrance and an exit,  ${\tt nparts} \ge 2$.

When tracking a test particle with initial conditions ($\sigma_x$, 0, $\sigma_y$, 0, 0, 0) for 5000 turns in both MADX and pyORBIT, the two codes produce very different results.
\begin{figure}%% [h]
    \centering
    \includegraphics[scale = 0.15]{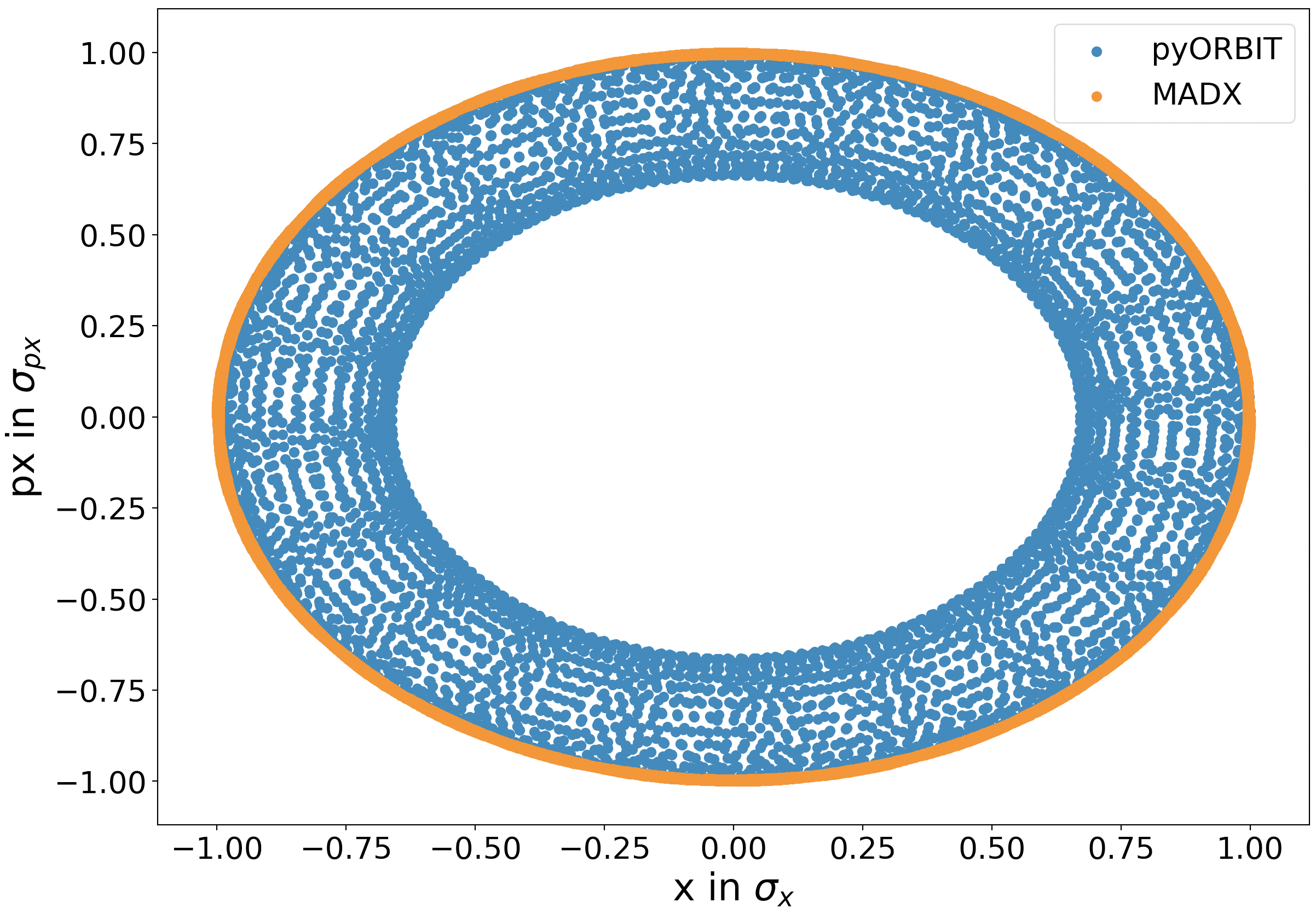}
    \includegraphics[scale = 0.15]{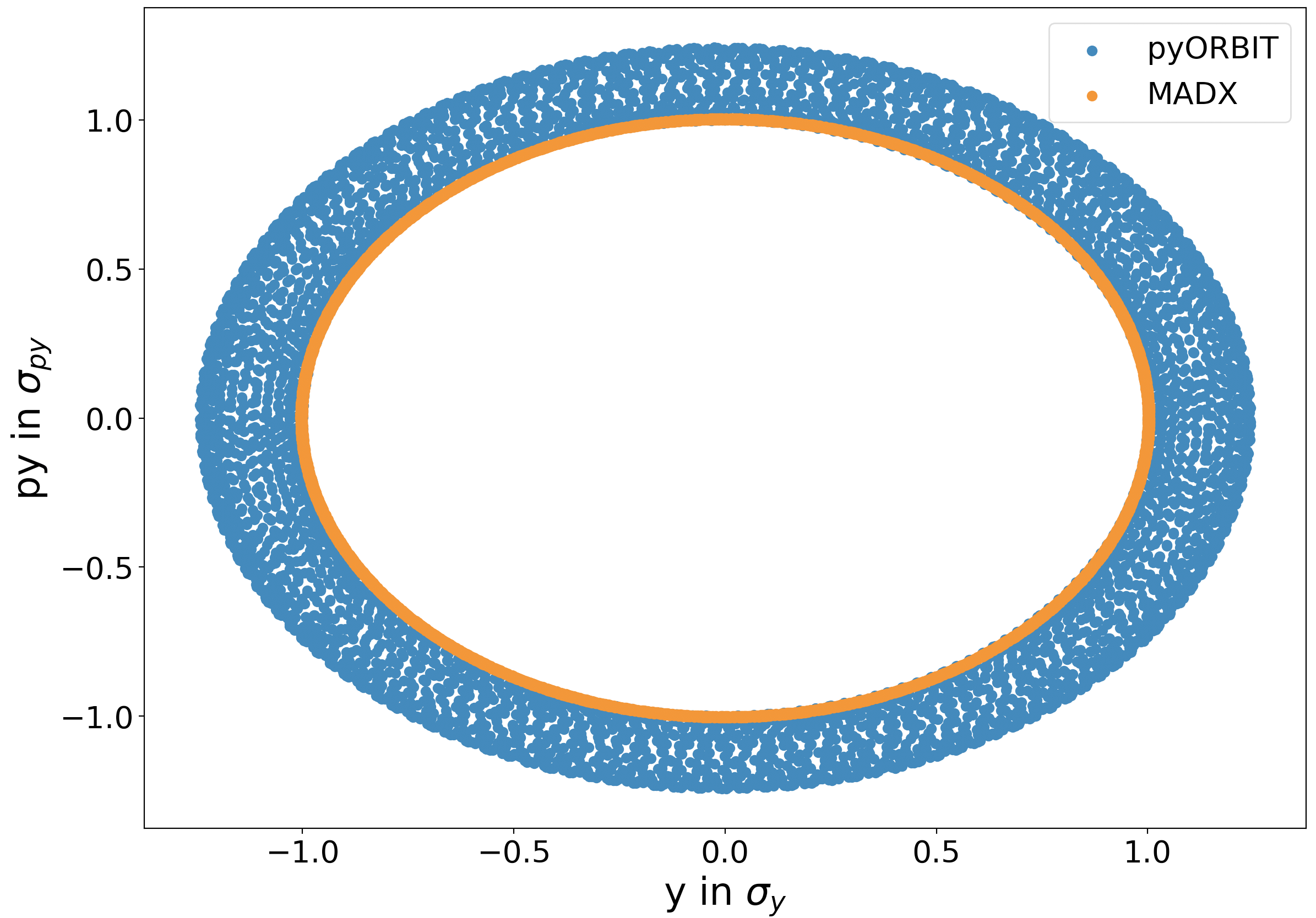}
    \caption{Single particle phase space diagram by MADX and pyORBIT}
    \label{fig:nonlinear_py}
\end{figure}

As shown in figure \ref{fig:nonlinear_py}, the trajectories in the $x$-$p_x$ and $y$-$p_y$ planes show differences on the order of $30\%$. The shape of phase space trajectory produced by {\tt pyORBIT} suggests the excitation of a resonance by presence of nonlinearities in {\tt pyORBIT}. After removing all sources of nonlinearity in {\tt pyORBIT}, we identified the {\tt Fringe In} and {tt Fringe Out} sections of the dipole and quadrupole element as the culprit. Their contributions are defined in the file 
{\tt py-orbit/src/teapot/teapotbase.cc}. \\
Removing these methods completely produced the figure \ref{fig:linear_py}.

\begin{figure} %% [h!]
    \centering
    \includegraphics[scale = 0.15]{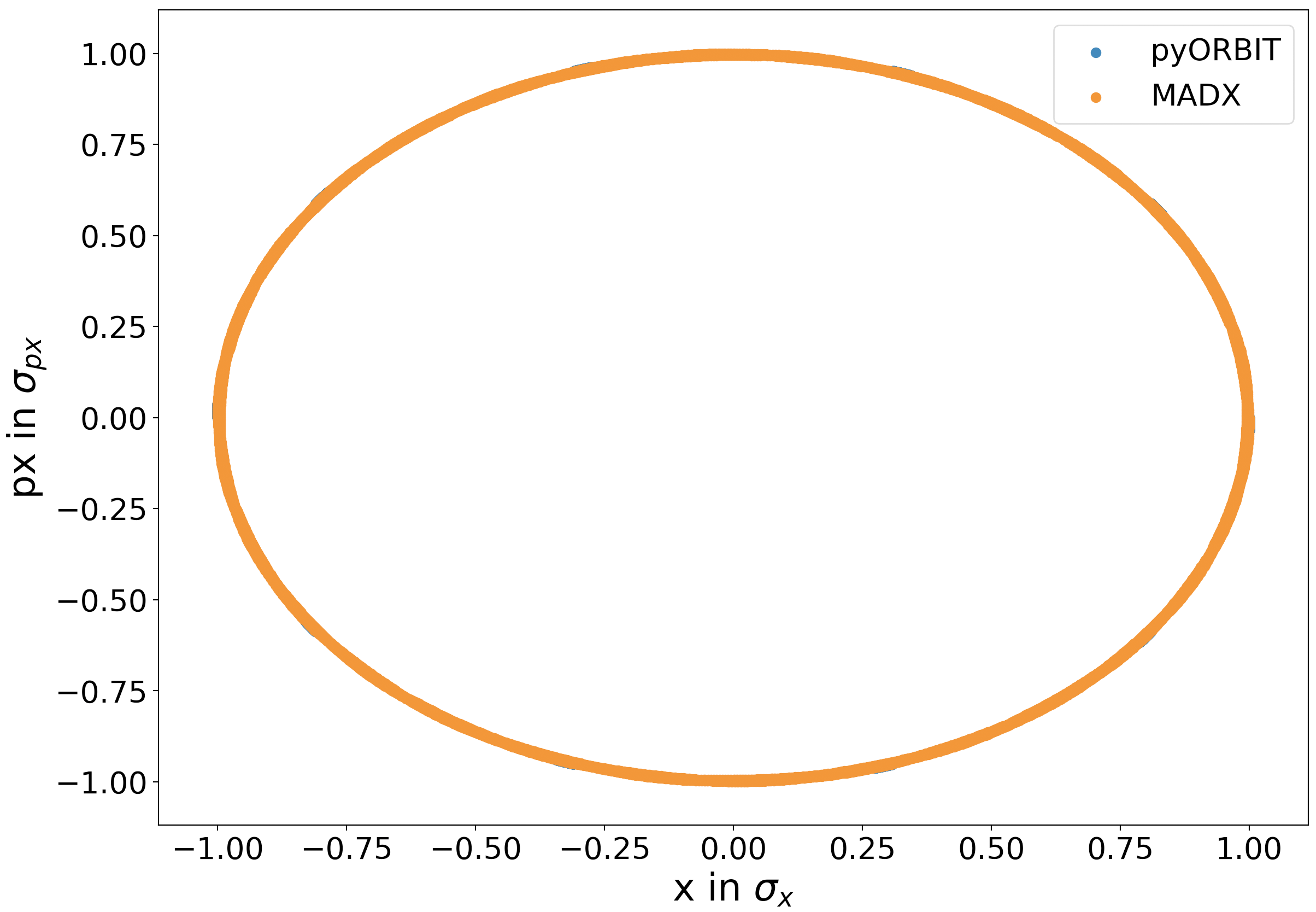}
    \includegraphics[scale = 0.15]{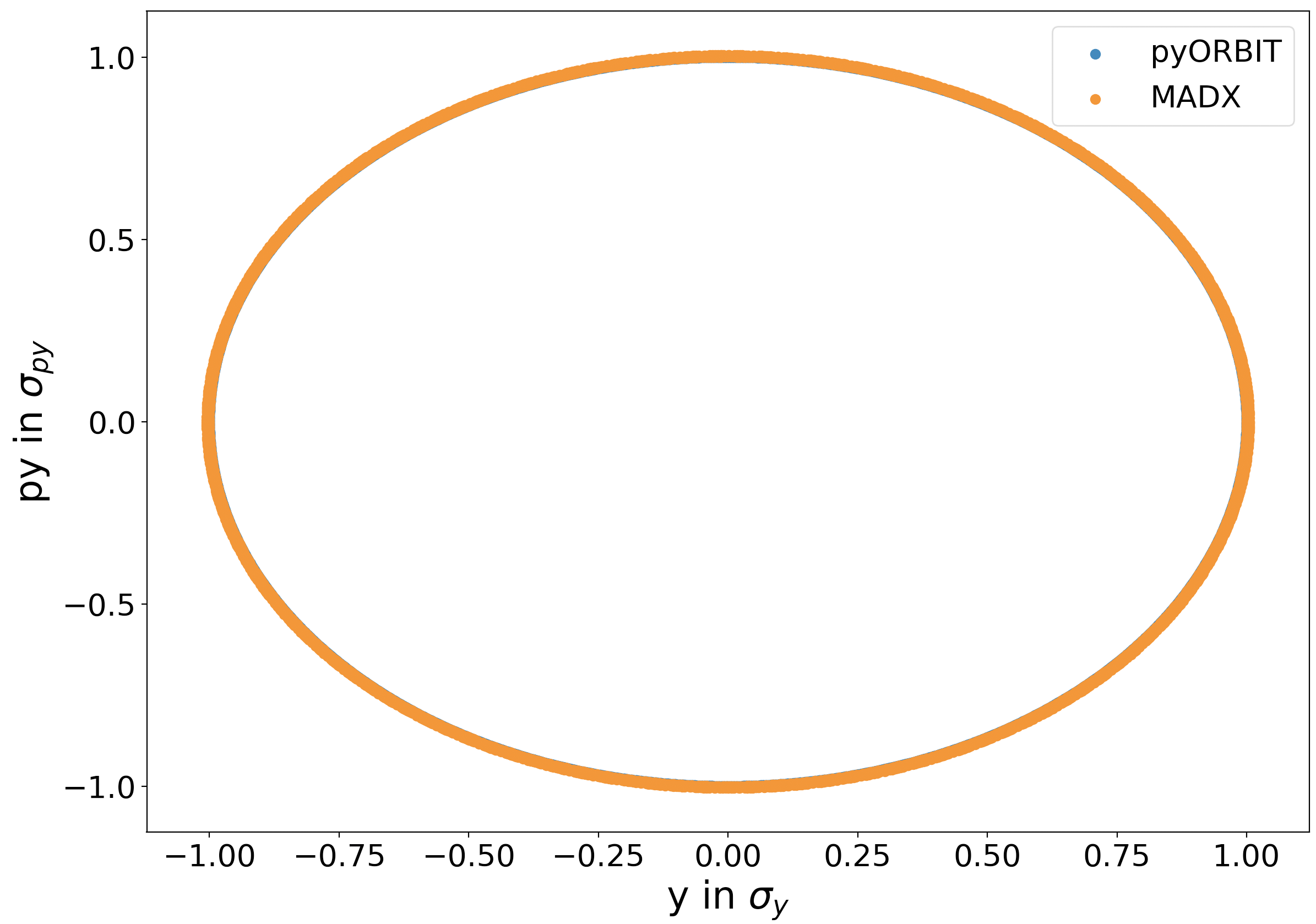}
    \caption{Single particle phase space diagram by MADX and pyORBIT after removing the magnet edge effect}
    \label{fig:linear_py}
\end{figure}

{ \tt MADX} does not consider nonlinear fringe effects; however, for small a ring like IOTA they may be significant. Additional tests would be required to assess the importance of nonlinear edge effects in IOTA.

\subsubsection{Other Changes}

A number of bugs in the {\tt pyORBIT} code were found and corrected. They are listed here. 

\begin{itemize}
\item In the IOTA sequence generated by {\tt MADX}, the strength of solenoids is set to the default value 0; 
this triggered a division by 0 error in {\tt pyORBIT}. 
As a workaround,  the strength of such solenoids can be  manually set to a very small value like $10^{-10}$.

\item In the file {\tt py-orbit/py/orbit/teapot/teapot.py}, the method {\tt  initialize()} of class
  {\tt ApertureTEAPOT} initializes the {\tt Aperture} object with\\
  {\tt Aperture(shape, dim[0], dim[1], 0.0, 0.0)},\\ however the {\tt Aperture} object is created by {\tt Aperture(int shape, double a, double b, double c, double d, double pos)} so a total of 6 parameters are needed. A dummy value 0.0 was added to fix this. 

\item Also, the treatment of apertures in {\tt pyORBIT} still has issues with sequence files generated by MADX. The {\tt pyORBIT} parser looks for elements of type {\tt Aperture} in the sequence; however, in the IOTA lattice an aperture can be defined using a marker with attribute "aperture". The current workaround is to remove all apertures defined in the sequence and add them back later using python code.

\item Other bugs fixed in {\tt pyORBIT} include three methods in the code
  \\ {\tt py-orbit/py/orbit/aperture/ApertureLatticeRangeModifications.py},\\
  which add apertures upstream and downstream of all elements, and one method in the file
  \\ {\tt py-orbit/src/orbit/ Aperture/Aperture.cc},\\
  which checks for particles intercepted at the aperture and moves them to the {\tt  lostbunch}.

\end{itemize}


\begin{thebibliography}{}
\bibi{pyORBIT} https://github.com/PyORBIT-Collaboration/py-orbit
\bibi{Wangler_91}T. Wangler, {\em Emittance growth from space charge forces}, Los Alamos report
LA-UR-91-3287
\bibi{Reiser}M. Reiser, {\em Theory and Design of Charged Particle Beams}, Wiley, New York
  \bibi{Danilov} V. ~Danilov and S.~Nagaitsev, {\em Nonlinear Accelerator Lattices with One and Two Analytic Invariants}, Phys. Rev. ST-AB, {\bf 13},  084002 (2010)
  \bibi{Antipov}S.~Antipov et al, {\em IOTA (Integrable Optics Test Accelerator): Facility and Experimental Beam Physics Program}, JINST {\bf 12}, T03002 (2017)
\bibi{Hoff_2014} I. Hoffmann and O. Boine-Frankenheim, {\em Grid dependent noise and entropy growth in anisotropic 3d particle-in-cell simulation of high intensity beams}, Phys. Rev. ST-AB, {\bf 17}, 124201 (2014)
\bibi{Kesting_2015} F. Kesting and G. Franchetti, {\em Propagation of numerical noise in particle-in-cell tracking},
Phys. Rev. ST-AB, {\bf 18}, 114201 (2015)
\bibi{Sen_unpub} T. Sen, unpublished notes
\bibi{Schmidt_2014}  F. Schmidt et al, {\em Code bench-marking for long-term tracking and adaptive algorithms},
  Proceedings of HB2016,, p 357 (2016)
\bibi{Syphers}D. Edwards and M.J. Syphers, Section 3.4.3 in {\em An Introduction to The Physics of High Energy Accelerators}, Wiley, New York
\bibitem{dip}
Karl L. Brown. {\em A First- and Second-Order Matrix Theory for the Design of Beam Transport Systems and Charged Particle Spectrometers}. SLAC Report-75

  \end{thebibliography}
\end{document}